\documentclass[10pt,aps,roupedaddress,kluwer]{revtex4}

\usepackage[latin5]{inputenc}
\usepackage{color}
\usepackage{latexsym}

\usepackage{graphicx}

\usepackage{dcolumn}

\input epsf

\begin{document}

\input psfig.sty

\def\lesssim{\mathrel{\hbox{\rlap{\hbox
{\lower4pt\hbox{$\sim$}}}\hbox{$<$}}}}
 
\def\gtrsim{\mathrel{\hbox{\rlap{\hbox
{\lower4pt\hbox{$\sim$}}}\hbox{$>$}}}}

\title{Code Development of Three-Dimensional General Relativistic 
 Hydrodynamics with AMR(Adaptive-Mesh Refinement) 
and Results From Special and General Relativistic Hydrodynamic}

\author{Orhan D\"{o}nmez}

\address{Nigde University Faculty of Art and Science, 
Physics Department, Nigde, Turkey} 

\date{\today}

\begin{abstract}
In this paper, the general procedure to solve the General Relativistic
Hydrodynamical(GRH) equations with Adaptive-Mesh Refinement (AMR) is 
presented. In order to achieve, the GRH 
equations are written in the conservation 
form to exploit their hyperbolic character. The numerical solutions of
general relativistic hydrodynamic equations are done by High
Resolution Shock Capturing schemes (HRSC), specifically designed to
solve non-linear hyperbolic systems of conservation laws. These
schemes depend on the characteristic information of the system. 
The Marquina fluxes with MUSCL left and right states are used to solve GRH
equations. First, different
test problems with uniform and AMR grids on the special relativistic
hydrodynamics equations are carried out to verify the second order
convergence  of the code in $1D$, $2D$ and $3D$. Results from uniform and AMR 
grid are compared. It is found that adaptive grid does a better job when 
the number of resolution is increased.  Second, the general relativistic 
hydrodynamical 
equations are tested using two different test problems which are Geodesic 
flow and Circular motion of particle In order to this, the flux 
part of GRH equations is coupled with source part using Strang splitting. 
The coupling of the GRH equations is carried out in a treatment which 
gives second order accurate solutions in space and time.
\end{abstract}



\maketitle




\section{Introduction}

Numerical relativity in astrophysics is becoming an exciting
area. The general relativistic observational data requires integration
of general relativistic hydrodynamical equations with  traditional
tools which need  high computational power. There are different numerical
tools to solve  the relativistic equations, such as general
relativistic hydrodynamics,
magnetohydrodynamics, and nuclear astrophysics. 

Relativity is necessary for describing astrophysical phenomena
involving compact objects. Some of  these phenomena are X-ray
binary systems, coalescing compact binaries, black hole formations,
gamma-ray burst, core collapse supernovae, pulsars, microquasars and
active galactic nuclei. The general relativistic effects must be
considered when strong gravitational fields are encountered, such as
near compact binaries. The significant gravitational wave signal
produced by some of these phenomena can also be understood in the 
framework of the general relativistic theory.

Numerical relativity in astrophysics is becoming more exiting area of
research due to the large amount of data taken from high energy
astronomical phenomena. Accordingly, the numerical relativistic 
astrophysics may
help to understand some of the observations in high energy  and
gravitational wave astrophysics. In order to take advantage from
observational data, relativistic equations must be  solved  to
understand phenomena in the strong relativistic regions. The numerical
solution of strong field astrophysical phenomena requires integration
of the Einstein theory  with other tools of astrophysics which are
relativistic hydrodynamics and nuclear astrophysics. The Einstein
theory describes the dynamics of the space time with a set of the
most complicate equations. It is highly nonlinear and thousands of
terms are coupled together in these equations.

The main issues in General Relativistic Hydrodynamic (GRH) are the
numerical modeling of flows with large Lorentz factors (strong
gravitational regions) and strong shock waves. An accurate description
of relativistic flows  with strong shocks is needed for study of
important problems in astrophysics. 
General relativistic effects caused by 
the presence of strong gravitational fields appear  to be connected to
extremely fast flows in different astrophysical scenarios, such as
accreting of compact objects, stellar collapse, and coalescing compact
binaries (Banyuls \emph{et al.}, 1997). In the last ten years, much 
effort has been addressed
to develop accurate numerical algorithms to solve the equations of
general relativistic hydrodynamics in extreme conditions. The main
conclusion that has emerged is that modern algorithms exploiting the
hyperbolic and conservative character of the equations are
more accurate at describing relativistic flows than
traditional finite difference upwind techniques with artificial
viscosity(Norman \emph{et al.}, 1986). To take
advantage of the conservation properties of the system, modern
algorithms are written in conservative form, in the sense that the
variation of mean values of the conserved quantities within the
numerical cells is given, in the absence of the sources, by the
fluxes across the cell boundaries. Furthermore, the hyperbolic
character of the system of equations allows one to treat physical
discontinuities appearing in the flow consistently (the
shock-capturing property)(Hirsch \emph{et al.}, 1992). Over the last few decades,
many High Resolution Shock Capturing (HRSC) methods have arisen in  
classical computational fluid  dynamics(Hirsch \emph{et al.}, 1992). Nowadays,
the numerical simulations employ these HRSC methods. These techniques produce
highly accurate numerical approximations in smooth regions of the flow
and capture the motion of unresolved steep gradients without creating
spurious oscillations. 

Our goal is to understand how thing formed in universe. Our strategy is the
application of known physics as it is dictated by the observations of the 
objects  whose origin and evolution we wish to comprehend, using and creating 
advanced computer techniques. Our tactics are to proceed based on 
three-dimensional numerical hydrodynamics using any source for computing 
methods wherever feasible, implementing the targeted physical processes in 
sequence, and proceeding to increasingly geometrically complex situations.

\section{FORMULATIONS}
\subsection{Description of Hydrodynamic Equations}
In this section, we give a general technical description of the Hydrodynamic
equation as they will be used in our code development and 
the analytic description of different problems.

The spacetime contains flow in 4-momentum. Each particle carries its
4-momentum vector  along its world line. The particles or
fluid have a 4-velocity $u^{\mu}(\mu = 0,1,2,3)$ at each point  which
is the $4$-velocity of an observer co-moving with the fluid at that
point. The unit of 4-velocity is unit of speed of light, c. At each
point there are three scalar functions  to define the
characteristic of the fluid. The first of these is the rest mass density 
$\rho(g/cm^3)$. The other scalar functions are the specific internal
energy $\epsilon(ergs/g)$ and the pressure $P$ in the units of $\rho
\epsilon$ which is the energy density. The pressure is connected to
the other variables by the  equation of state $P = P(\rho,
\epsilon)$. The quantity $\rho h$,  where $h$
is called the relativistic enthalpy, is the total inertia carrying
mass energy.

The time evolution of the fluid is governed by the General
Hydrodynamical equations. The first equation is the
normalization of the $4$-velocity, defined as

\begin{eqnarray} 
u^{\mu} u_{\mu} = -1.
\label{des 1}
\end{eqnarray}

\noindent
The GRH equations in  references
Font \emph{et al.}, 2000 and Donat \emph{et al.}, 1998, written in the standard
covariant form, consist  of the local conservation laws of the
stress-energy tensor $T^{\mu \nu }$  and the matter current density $
J^\mu$:

\begin{eqnarray}
\bigtriangledown_\mu T^{\mu \nu} = 0 ,\;\;\;\;\;
\bigtriangledown_\mu J^\mu = 0.
\label{covariant derivative}
\end{eqnarray}

\noindent
Greek indices run from 0 to 3, Latin indices
from 1 to 3, and units in which the speed of light $c=1$ are
used. $\bigtriangledown_\mu$
stands for the covariant derivative with respect to the 4-metric of the
underlying spacetime, $g_{\mu \nu}$.

In order to  quantify the  fluid or particles which are moving in
their world line, the stress-energy tensor $T^{\mu \nu}$ is
defined. The stress-energy tensor exists at each event in
spacetime. The stress-energy density is a quantity which contains
information about energy density, momentum density and stress as
measured by any and all observers at that event. The stress-energy
tensor for an imperfect fluid is defined by Misner \emph{et al.} as 

\begin{eqnarray} 
T^{\mu \nu} = \rho(1+\epsilon) u^{\mu} u^{\nu} + (P - \zeta
\theta)h^{\mu \nu} - \nonumber \\ 
2 \eta \sigma^{\mu \nu} + q^{\mu}u^{\nu} + q^{\nu}u^{\mu},
\label{des 4}
\end{eqnarray}

\noindent
where the scalar functions $\eta$ and $\zeta$ are the shear and bulk
viscosities, respectively. This is the stress-energy tensor for a
viscous fluid with heat conduction. The scalar $\theta$ is defined as 

\begin{eqnarray} 
\theta = u^{\alpha}_{\;\; ; \alpha}.
\label{des 5}
\end{eqnarray}

\noindent
It describes the divergence of the fluid world lines. The symmetric,
trace-free, and spatial shear tensor $\sigma^{\mu \nu}$ is defined by

\begin{eqnarray} 
\sigma^{\mu \nu} = \frac{1}{2}(u^{\mu}_{\;\; ; \gamma} h^{\gamma \nu}
+ u^{\nu}_{\;\; ; \gamma}h^{\gamma \mu}) - \frac{1}{3}\theta h^{\mu
\nu},
\label{des 6}
\end{eqnarray}

\noindent
where $h_{\gamma \mu} = u_{\gamma} u_{\mu} + g_{\gamma \mu}$ is the
spatial projection tensor. The vector $q^{\mu}$ is the energy flux
vector. 

Eqs.(\ref{des 4} - \ref{des 6}) describe the motion of imperfect
fluid flow for a fixed metric in complete generality. Entropy 
for imperfect fluid is influenced by viscosity and heat flow. 

Defining the characteristic waves of the general
relativistic hydrodynamical equations is not trivial with imperfect
fluid stress-energy tensor. The viscosity and heat
conduction effects are neglected.  This defines the  perfect fluid
stress-energy tensor. This stress-energy tensor is used  to derive the
hydrodynamical equations. Using 
perfect fluid stress-energy tensor, we can solve some problems which
are  solved by the Newtonian hydrodynamics with viscosity, such as
those involving angular momentum transport and shock waves on an
accretion disk, etc. Entropy for 
perfect fluid is conserved along the fluid lines. The stress energy tensor
for a perfect fluid is given as

\begin{equation}
T^{\mu \nu} = \rho h u^\mu u^\nu + P g^{\mu \nu}.
\label{des 7}
\end{equation}

\noindent
A perfect fluid  is a fluid that moves through spacetime with a
4-velocity $u^{\mu}$ which may vary from event to event. It exhibits a
density of mass $\rho$ and isotropic pressure $P$ in the rest frame of
each fluid element. $h$ is the specific
enthalpy, defined as

\begin{equation} 
h = 1 + \epsilon +\frac{P}{\rho}.
\label{hdot}
\end{equation}

\noindent
Here  $\epsilon$ is the specific internal energy. The equation of
state might have  the 
functional form $P = P(\rho, \epsilon)$. The perfect
gas equation of state, 

\begin{equation}
P = (\Gamma -1 ) \rho \epsilon,
\label{flux split21}
\end{equation}

\noindent
is such a functional form.

The conservation laws in the form given in Eq.(\ref{covariant
derivative}) are not suitable for 
the use in advanced numerical schemes. In order to carry out numerical
hydrodynamic evolutions such as those reported in Font \emph{et al.}, 2000, 
and to
use  HRSC methods, the hydrodynamic equations after the 3+1 split
must be written as a hyperbolic system of first order flux
conservative equations. The Eq.(\ref{covariant derivative}) is written in
terms of coordinate derivatives, using the coordinates ($x^0 = t, x^1,
x^2, x^3$). Eq.(\ref{covariant derivative}) is projected onto the
basis $\lbrace n^\mu, (\frac{\partial}{\partial x^i})^\mu \rbrace$,
where $n^\mu$ is a unit timelike vector normal to a given
hypersurface. After a straightforward calculation,
we get (see ref.Font \emph{et al.}, 2000),

\begin{equation}
\partial_t \vec{U} + \partial_i \vec{F}^i = \vec{S}, 
\label{desired equation}
\end{equation} 

\noindent
where $\partial_t = \partial / \partial t$ and $\partial_i = \partial
/ \partial x^i$. This basic step serves to identify the
set of unknowns, the vector of conserved quantities $\vec{U}$, and
their corresponding fluxes $\vec{F}(\vec{U})$. With the
equations in conservation form, almost every high
resolution method devised to solve hyperbolic systems of conservation
laws can be extended to GRH.

The evolved state vector $\vec{U}$ consists of  the conservative
variables $(D, S_j, \tau)$ which are conserved variables for density,
momentum and energy respectively; in terms of the
primitive variables $(\rho, v^i, \epsilon)$, this becomes(Font \emph{et al.},
 2000)

\begin{equation}
\vec{U} = \left( 
\begin{array}{c} 
D \\
S_j \\
\tau 
\end{array} 
\right) = \left( 
\begin{array}{c}
\sqrt{\gamma} W \rho \\
\sqrt{\gamma}\rho h W^2 v_j \\
\sqrt{\gamma} (\rho h W^2 - P - W \rho) 
\end{array} 
\right). 
\label{matrix form of conserved quantities}
\end{equation}

\noindent
Here $\gamma$ is the determinant of the 3-metric $\gamma_{ij}$, $v_j$
is the fluid 3-velocity, and W is the Lorentz factor,

\begin{equation}
W = \alpha u^0 = (1 - \gamma_{ij} v^i v^j)^{-1/2}. 
\label{Wdot1}
\end{equation}

\noindent 
The flux vectors $\vec{F^i}$ are given by(Font \emph{et al.}, 2000)

\begin{equation}
\vec{F}^i =  \left( \begin{array}{c}
\alpha (v^i - \frac{1}{\alpha} \beta^i) D \\
\alpha \lbrace (v^i - \frac{1}{\alpha} \beta^i) S_j + \sqrt{\gamma} P
\delta ^{i}_{j} \rbrace    \\
\alpha \lbrace (v^i - \frac{1}{\alpha} \beta^i) \tau + \sqrt{\gamma}
v^i P \rbrace \end{array} \right). 
\label{matrix form of Flux vector}
\end{equation}

\noindent
The spatial components of the 4-velocity $u^i$ are related to the
3-velocity by the following formula: $u^i = W (v^i - \beta^i /
\alpha)$. $\alpha$ 
and $\beta^i$ are the lapse function and the shift vector of the
spacetime respectively. The source vector $\vec{S}$ is given by
(Font \emph{et al.}, 2000)

\begin{equation}
\vec{S} =  \left( \begin{array}{c}
0 \\
\alpha \sqrt{\gamma} T^{\mu \nu} g_{\nu \sigma} \Gamma^{\sigma}_{\mu j}  \\
\alpha \sqrt{\gamma} (T^{\mu 0} \partial_{\mu} \alpha - \alpha T^{\mu
\nu} \Gamma^{0}_{\mu \nu}) \end{array} \right), 
\label{matrix form of source vector}
\end{equation}

\noindent
where $\Gamma^{\alpha}_{\mu \nu}$ is the 4-dimensional Christoffel symbol

\begin{equation}
\Gamma^{\alpha}_{\mu \nu} = \frac{1}{2} g^{\alpha \beta}
(\partial_\mu g_{\nu \beta} + \partial_\nu g_{\mu \beta} -
\partial_\beta g_{\mu \nu}).
\label{Christoffer symbol}
\end{equation}


\subsection{Spectral Decomposition and Characteristic Fields}
\label{Spectral Decomposition and Characteristic Fields}

The use of HRSC  schemes requires the
spectral decomposition of the Jacobian matrix of the
system, $\partial \vec{F^i} / \partial \vec{U}$. 
The spectral decomposition of the Jacobian matrices of the GRH 
 equations with a general equation of state
was reported in Font \emph{et al.}, 2000. They display the 
spectral decomposition, valid for a generic spatial metric, 
in the x-direction, $(\partial \vec{F}^x / \partial \vec{U})$;
permutation of the indices yields the other two directions.   

They started the solution by considering an equation of state in which the
pressure $P$ is a function of $\rho$ and $\epsilon$, $P = P(\rho,
\epsilon)$. The relativistic speed of sound in the fluid $C_s$ is
given by(Font \emph{et al.}, 2000)

\begin{equation}
C_{s}^2 = \left. \frac{\partial P}{\partial E} \right|_S. =  \frac{\chi}{h} + \frac{P
\kappa}{\rho^2 h},   
\label{cdot}
\end{equation}

\noindent
where $\chi = \partial P / \partial \rho\vert_\epsilon$, $\kappa = \partial
P / \partial \epsilon\vert_\rho$, $S$ is the entropy per particle, and
$E = \rho + \rho \epsilon$ is the total rest energy density.


A complete set of the right eigenvectors $[\vec{r}_i]$ and  corresponding
eigenvalues $\lambda_i$  along the $x$-direction obeys

\begin{eqnarray}
\left[ \frac{\partial \vec{F}^x}{\partial \vec{U}} \right] \left[
\vec{r_i} \right] = 
\lambda_i \left[ \vec{r_i} \right], \; \; \; i= 1,...,5.
\label{eigenvalue-vector equation}
\end{eqnarray}

\noindent
The solution contains triply degenerate eigenvalues(Font \emph{et al.}, 2000),

\begin{eqnarray}
\lambda_1^x = \lambda_2^x = \lambda_3^x = \alpha v^x - \beta^x.
\label{eigenvalue123}
\end{eqnarray}

\noindent
The other eigenvalues are given by(Font \emph{et al.}, 2000)

\begin{eqnarray}
\lambda_{\pm}^x = \frac{\alpha}{1-v^2 C_s^2} \Big\{v^x (1-C_s^2) \pm
\nonumber \\
\sqrt{C_s^2 (1-v^2) [\gamma^{xx} (1-v^2 C_s^2) - v^x v^x (1 - C_s^2)]}
\Big\} -\beta^x.
\label{eigenvalue plus minus}
\end{eqnarray}

\noindent
A set of linearly independent right eigenvectors spanning  this space
is given by(Font \emph{et al.}, 2000)

\begin{eqnarray}
\vec{r}_1^x = \left[ \frac{\kappa}{h W (\kappa - \rho C_s^{2})}, v_x, v_y,
v_z, 1 - \frac{\kappa}{h W (\kappa - \rho C_s^{2})} \right]^T,
\label{eigenvector1}
\end{eqnarray}

\begin{eqnarray}
\vec{r}_2^x = \bigg[ W v_y, h (\gamma_{xy} + 2 W^2 v_x v_y), h (\gamma_{yy} +
2 W^2 v_y v_y),  \nonumber  \\ 
 h (\gamma_{yz} + 2 W^2 v_y v_z), v_y W (2 W h -1)\bigg]^T,
\label{eigenvector2}
\end{eqnarray}

\begin{eqnarray}
\vec{r}_3^x = \bigg[ W v_z, h (\gamma_{xz} + 2 W^2 v_x v_z), h (\gamma_{yz} +
2 W^2 v_y v_z), \nonumber \\ 
h (\gamma_{zz} + 2 W^2 v_z v_z), v_z W (2 W h -1) \bigg]^T,
\label{eigenvector3}
\end{eqnarray}

\noindent
and

\begin{eqnarray}
\vec{r}_{\pm}^x = \bigg[1, h W (v_x - \frac{v^x -(\lambda_{\pm} +
\beta^x)/\alpha}{\gamma^{xx} - v^x (\lambda_{\pm} + \beta^x)/\alpha}), h W
v_y, h W v_z, \nonumber \\ 
\frac{h W (\gamma^{xx} -v^x v^x)}{\gamma^{xx} -v^x (\lambda_{\pm} +
\beta^x)/\alpha} - 1 \bigg]^T. 
\label{eigenvector plus minus}
\end{eqnarray}

\noindent
Here, the superscript T denotes the transpose.

Now, the eigenvalues and eigenvectors are computed in the $y$ and
$z$-directions using the information in the $x$-direction.
Let ${\bf M}^1(U)$ be the matrix of the right eigenvectors, that is, the
matrix having as columns the right eigenvectors with the standard
ordering(Banyuls \emph{et al.}, 1997), ${\bf M}^1 = (\vec{r}_-, \vec{r}_1, \vec{r}_2,
\vec{r}_3, \vec{r}_+)$.  To obtain the spectral decomposition in the
other spatial directions $q = 2,3 (\equiv y,z)$, it is enough to take
into account the following symmetry relations:

\newcounter{abeann}
\begin{list}%
{\arabic{abeann})}{\usecounter{abeann}
	\setlength{\rightmargin}{\leftmargin}}

\item The eigenvalues are easily obtained by carrying out in equations
(\ref{eigenvalue123}) and (\ref{eigenvalue plus minus}) a simple
substitution of indices $x$ by $q$. 

\item The matrix $M^1$ in the $x$-direction is permuted according to 
${\bf M^q} = {\cal P}_{q1}({\bf M}^1)$,
where the two operators ${\cal P}_{q1}$ act on their arguments by the
following sequential operations:

\begin{itemize}

\item to permute the second and $(q + 1)$th rows,

\item to interchange indices $1 \equiv x$ with $q$.

\end{itemize}

\end{list}

From Ibanez \emph{et al.}, 1998, the left
eigenvectors in the $x$-direction are :

\begin{eqnarray}                                                      
{\bf l}_{0,1} = {\displaystyle{\frac{W}{{\cal K} - 1}}} 
\left[ \begin{array}{c}
h - W
\\  \\
W v^x
\\  \\
W v^y
\\  \\
W v^z
\\  \\
- W
\end{array} \right]
\end{eqnarray}

\begin{eqnarray}                                                     
{\bf l}_{0,2} = {\displaystyle{\frac{1}{h \xi^x}}}
\left[ \begin{array}{c}
- \gamma_{zz} v_y + \gamma_{yz} v_z
\\  \\
v^x (\gamma_{zz} v_y - \gamma_{yz} v_z)
\\  \\
\gamma_{zz} (1 - v_x v^x) + \gamma_{xz} v_z v^x 
\\  \\
- \gamma_{yz} (1 - v_x v^x) - \gamma_{xz} v_y v^x 
\\  \\
- \gamma_{zz} v_y + \gamma_{yz} v_z
\end{array} \right]
\end{eqnarray}                                                               

\begin{eqnarray} 
{\bf l}_{0,3} = {\displaystyle{\frac{1}{h \xi^x}}}
\left[ \begin{array}{c}
- \gamma_{yy} v_z + \gamma_{zy} v_y
\\  \\
v^x (\gamma_{yy} v_z - \gamma_{zy} v_y)
\\  \\
- \gamma_{zy} (1 - v_x v^x) - \gamma_{xy} v_z v^x 
\\  \\
\gamma_{yy} (1 - v_x v^x) + \gamma_{xy} v_y v^x 
\\  \\
- \gamma_{yy} v_z + \gamma_{zy} v_y
\end{array} \right]                                                             
\end{eqnarray}

\begin{eqnarray} 
{\bf l}_{\mp} = ({\pm} 1){\displaystyle{\frac{h^2}{\Delta^x}}} 
\left[ \begin{array}{c} 
h W {\cal V}^x_{\pm} \xi^x + (\pm \frac{\Delta}{h^2}) l_{\mp}^{(5)} 
\\  \\
\Gamma_{xx} (1 - {\cal K} {\tilde {\cal A}}^x_{\pm}) +
(2 {\cal K} - 1){\cal V}^x_{\pm} (W^2 v^x \xi^x - \Gamma_{xx} v^x)
\\  \\
\Gamma_{xy} (1 - {\cal K} {\tilde {\cal A}}^x_{\pm}) +
(2 {\cal K} - 1) {\cal V}^x_{\pm} (W^2 v^y \xi^x - \Gamma_{xy} v^x)
\\  \\
\Gamma_{xz} (1 - {\cal K} {\tilde {\cal A}}^x_{\pm}) +
(2 {\cal K} - 1) {\cal V}^x_{\pm} (W^2 v^z \xi^x - \Gamma_{xz} v^x)
\\  \\
(1 - {\cal K})[- \gamma v^x + {\cal V}^x_{\pm} (W^2 \xi^x - \Gamma_{xx})]
- {\cal K} W^2 {\cal V}^x_{\pm} \xi^x 
\end{array} \right]
\end{eqnarray}   


\noindent
The variables used for the left eigenvectors in the $x$-direction are defined as follows:

\begin{eqnarray*}
\Lambda_{\pm}^x \equiv {\tilde{\lambda}}_{\pm} + \tilde{\beta}^x
\,\,\,\,\,,\,\,\,\,\,
\tilde{\lambda} \equiv \lambda/\alpha
\,\,\,\,\,,\,\,\,\,\, 
\tilde{\beta}^x \equiv \beta^x/\alpha 
\end{eqnarray*}

\begin{eqnarray*}
{\cal K} \equiv {\displaystyle{\frac{\tilde{\kappa}}
{\tilde{\kappa}-c_s^2}}} \,\,\,\,\,,\,\,\,\,\,
\tilde{\kappa}\equiv \kappa/\rho 
\end{eqnarray*}

\begin{eqnarray*}
{\cal C}^x_{\pm} \equiv v_x - {\cal V}^x_{\pm} \,\,\,,\,\,\,
{\cal V}^x_{\pm} \equiv {\displaystyle{\frac{v^x - \Lambda_{\pm}^x}
{\gamma^{xx} - v^x \Lambda_{\pm}^x}}} 
\end{eqnarray*}

\begin{eqnarray*}
{\tilde {\cal A}}^x_{\pm} \equiv 
{\displaystyle{\frac{\gamma^{xx} - v^x v^x}
{\gamma^{xx} - v^x {\Lambda}^x_{\pm}}}}
\end{eqnarray*}

\begin{eqnarray*}
1 - {\tilde {\cal A}}^x_{\pm} = v^x {\cal V}^x_{\pm} \,\,\,,\,\,\,
{\tilde {\cal A}}^x_{\pm} - {\tilde {\cal A}}^x_{\mp} =
v^x ({\cal C}^x_{\pm} - {\cal C}^x_{\mp}) 
\end{eqnarray*}

\begin{eqnarray*}
({\cal C}^x_{\pm} - {\cal C}^x_{\pm}) +
({\tilde {\cal A}}^x_{\mp} {\cal V}^x_{\pm} -
{\tilde {\cal A}}^x_{\pm} {\cal V}^x_{\mp}) = 0 
\end{eqnarray*}

\begin{eqnarray*}
\Delta^x \equiv h^3 W ({\cal K} - 1) ({\cal C}^x_{+} - {\cal C}^x_{-}
) \xi^x 
\end{eqnarray*}

\begin{eqnarray*}
\xi^x \equiv \Gamma_{xx}  - \gamma v^x v^x 
\end{eqnarray*}

\begin{eqnarray*}
\gamma \equiv det \gamma_{ij} = \gamma_{xx} \Gamma_{xx} +
\gamma_{xy} \Gamma_{xy} + \gamma_{xz} \Gamma_{xz}
\end{eqnarray*}

\begin{eqnarray*}
\Gamma_{xx} = \gamma_{yy} \gamma_{zz} - \gamma_{yz} \gamma_{zy} \\ 
\Gamma_{xy} = \gamma_{yz} \gamma_{zx} - \gamma_{yx} \gamma_{zz} \\
\Gamma_{xz} = \gamma_{yx} \gamma_{zy} - \gamma_{yy} \gamma_{zx}
\end{eqnarray*}

\begin{eqnarray*}
\Gamma_{xx} v_x + \Gamma_{xy} v_y + \Gamma_{xz} v_z= 
\gamma v^x 
\end{eqnarray*}

Now, the left eigenvectors of the Jacobian matrix are computed in the $y$
and $z$-directions using  the left eigenvectors in the $x$-direction. 
Let ${\bf L}^1(U)$ be the matrix of left eigenvectors, that is, the
matrix having as rows the left eigenvectors with the standard
ordering. The eigenvalues of the left eigenvectors are the same with the eigenvalues 
of the right eigenvectors. To obtain the spectral decomposition in the
other spatial directions $q = 2,3 (\equiv y,z)$, it is enough to take
into account the following symmetry relations:

\noindent
The matrix $M^1$ in the $x$-direction is permuted according to 
${\bf L^q} = {\cal P}_{q1}({\bf L}^1)$,
where the two operators ${\cal P}_{q1}$ act on their arguments by the
following sequential operations:

\begin{itemize}

\item to permute the second and $(q + 1)$th columns,

\item to interchange indices $1 \equiv x$ with $q$.

\end{itemize}


\section{NUMERICAL METHOD}

\subsection{Marquina Fluxes}
\label{Marquina Fluxes}

Approximate Riemann solver failures and their respective corrections
(usually adding a  artificial dissipation) have been studied in the
literature(Quirk, 1994). Motivated by the search for a robust and accurate
approximate Riemann solver that avoids these common failures, 
Shu \emph{et al.}, 1989 have proposed a numerical flux formula for scalar
equations. Marquina flux is generalization of flux formula in
Ref.Shu \emph{et al.}, 1989. In the scalar case and for
characteristic wave speeds which do not change sign at the given
numerical interface, Marquina's flux formula is identical to Roe's
flux(Toro, 1999).  Otherwise, scheme is more viscous, entropy
satisfying local Lax-Friedrichs scheme(Shu \emph{et al.}, 1989). The
combination of Roe and Lax-Friedrichs schemes is carried out in each
characteristic field after the local linearization and decoupling of
the system of equations. However, contrary to other schemes, the
Marquina's method is not based on any averaged intermediate state.

The Marquina fluxes(Font \emph{et al.}, 2000) with MUSCL left and right states are used to solve the 
3-D relativistic hydro equation. In Marquina's scheme there are no
Riemann solutions 
involved (exact or approximate) and there are no artificial
intermediate states constructed at each cell interface. 

To compute the Marquina fluxes,first,the sided local characteristic 
variables and fluxes are computed . For the left and right sides,
the characteristic variables are

\begin{equation}
w_{l}^{p} = L^{p}(U_{l}) \cdot U_l, \;\;\;\; w_{r}^{p} = L^{p}(U_{r}) \cdot U_r
\label{wl} 
\end{equation}

\noindent
and the characteristic fluxes are

\begin{equation}
\Phi_{l}^{p} =  L^{p}(U_{l}) \cdot F(U_l), \;\;\;\;\; \Phi_{r}^{p} =
L^{p}(U_{r}) \cdot F(U_r). 
\end{equation}

\noindent
where the number of conservative variables $p = 1..5$. $U_l$ and $U_r$
are conservative variables at the left and 
right sides, respectively. $L^{p}(U_{l})$ and $L^{p}(U_{r})$ are the left 
eigenvectors of the Jacobian matrice, $\partial F^i/\partial U$.

The left and right fluxes are defined depending on the velocities of 
the fluid for each specific grid zone. The prescription given in 
Ref.Donat \emph{et al.}, 1998 is
as follows.

For all conserved variables $p = 1, ..m$

if \emph{$\lambda_{p}(U)$} does not change sign in (if
( \emph{$\lambda_{p}(U_L)$ $\times$ $\lambda_{p}(U_R)$ $\geq$ 0))}, then 

$\;\;\;\;$ if \emph{$\lambda_{p}(U_l) > 0$} then

$\;\;\;\;\;\;\;\;\;\;\;$	$\Phi_{+}^{p} = \Phi_{l}^{p}$
	
$\;\;\;\;\;\;\;\;\;\;\;$	$\Phi_{-}^{p} = 0$

$\;\;\;\;$   else
	
$\;\;\;\;\;\;\;\;\;\;\;$	$\Phi_{+}^{p} = 0$
	
$\;\;\;\;\;\;\;\;\;\;\;$	$\Phi_{-}^{p} = \Phi_{r}^{p}$	

$\;\;\;\;$ end if

else

$\;\;\;\;\;\;\;\;\;\;\;$   $\alpha_{p} =  \max_{U \epsilon \Gamma(U_l,
U_r)} | \lambda_p(U) | $

$\;\;\;\;\;\;\;\;\;\;\;$ $\Phi_{+}^{p} = 0.5 ( \Phi_{l}^{p} +
\alpha_{k} w_{l}^{p})$

$\;\;\;\;\;\;\;\;\;\;\;$ $\Phi_{-}^{p} = 0.5 ( \Phi_{r}^{p} +
\alpha_{p} w_{r}^{p})$ 

end if

\noindent
where $\lambda_p$ is an eigenvalue of the Jacobian matrix and,
\begin{eqnarray}
\alpha_k = \max \{ |\lambda_{p} (U_l)|, |\lambda_{p} (U_r)| \}.
\label{alpha k}
\end{eqnarray}

\noindent
The numerical flux that corresponds to the cell interface
separating the states $U_l$ and $U_r$ is then given by 
Ref.Donat \emph{et al.}, 1998:

\begin{eqnarray}
F^{M} (U_l, U_r) = \sum_{p=1}^{m} ( \Phi_{+}^p r^p (U_l) +
\Phi_{-}^p r^p (U_r)  ).
\label{marquina flux}
\end{eqnarray}

Marquina's scheme can be  interpreted as a characteristic-based
scheme that avoids the use of an averaged intermediate state to
perform the transformation to the local characteristic fields. 

In carrying out Marquina's scheme, we have to compute intermediate
states and the Jacobian matrix of the states at each cell interface. So we
need to know the left and right states, $U_L$ and  $U_R$, at each
interface. To construct the 
second-order scheme, the MUSCL left and right states are used,

\begin{eqnarray}
u^{L}_{i} = u_{i}(0) = u_{i} - \frac{1}{2} \Delta_i \nonumber \\
u^{R}_{i} = u_{i}(\Delta x) = u_{i} + \frac{1}{2} \Delta_i .
\label{left and right MUSCL}
\end{eqnarray}

To avoid the appearance of oscillations around discontinuities in
MUSCL-type schemes, slope limiters in the reconstruction stage(Hirsch 
\emph{et al.}, 1992, Toro, 1999) are used .A common TVD limiter is 
based on the minmod function
(Hirsch \emph{et al.}, 1992). The standard minmod slope
provides the desired second-order accuracy for smooth solutions,
while still satisfying the $TVD$ property. Minmod slope function is given as 

\begin{eqnarray}
\Delta_i = \rm{minmod}(U_i - U_{i-1}, U_{i+1} - U_i),
\label{minmod1}
\end{eqnarray}

\noindent
where the minmod function of two arguments is defined by:

\begin{eqnarray}
 \rm{minmod}(a,b) = \left\{ \begin{array}{lll}
a & \mbox{if $|a| < |b|$ and $ab>0$} \\
b & \mbox{if $|b| < |a|$ and $ab>0$} \\
0 & \mbox{if $ab \leq 0$} .
		\end{array}
	\right. 
\label{minmod2}
\end{eqnarray}


\subsection{Strang Splitting}
\label{The Strang Splitting}

HRSC methods are applicable  without source term.
We now turn our attention to handling the source and flux terms. To solve 
the general relativistic hydrodynamical equations
numerically Strang splitting is used, in which Eq.(\ref{desired equation})
is split into two parts. The first is given by
the flux terms(Bona \emph{et al.}, 1998)

\begin{equation}
\frac{\partial \vec{U}}{\partial t} + \frac{\partial \vec{F}^x}{\partial
x} + \frac{\partial \vec{F}^y}{\partial y} + \frac{\partial
\vec{F}^z}{\partial z} = 0,
\label{Flux Part of Equation}
\end{equation}

\noindent
The evolution operator of Eq.(\ref{Flux Part of
Equation}) can be described by an $F$, so the solution of 
Eq.(\ref{Flux Part of Equation}) becomes

\begin{equation}
U(t+\Delta t) = F U(t)
\label{part 1}
\end{equation}

\noindent
The second has the source terms

\begin{equation}
\frac{\partial \vec{U}}{\partial t} = \vec{S}.
\label{source}
\end{equation}

\noindent
With the evolution operator for  Eq.(\ref{source}),
 $S$, the solution of Eq.(\ref{source})  is 

\begin{equation}
U(t+\Delta t) = S U(t)
\label{part 2}
\end{equation}

\noindent
In the numerical solution, this splitting is performed by a combination of both
flux and source operators. 
Denoting by $U(\Delta t)$ the numerical
evolution operator  for the system in a single time step,
the evolution operator for solution(Toro, 1999) is written as:

\begin{equation}
U(t+\Delta t) = S(\frac{\Delta t}{2}) F(\Delta t) S(\frac{\Delta t}{2})U(t).
\label{U operator}
\end{equation}

\noindent
The resulting update of $U(t+\Delta t)$ is second-order accurate in
space and time if and only if both the operators $S$ and $F$ are
$O(\Delta x^2, \Delta t^2)$.

\noindent
This kind of  splitting allows easy implementation of the
different numerical treatments of the principal part of the system
without worrying about the sources of the equations. It also allows
choosing the best scheme for each type of problem.


\subsection{Updating the Equations in Time}
\label{Updating the Equations in Time}

The HRSC methods will be used to solve
the hydrodynamical equations. This requires that the
hydrodynamical equations are written  in a flux form.

Solution of the relativistic hydrodynamical  equations using MUSCL left and
right states with Marquina fluxes  gives second-order
accurate solutions in space. In order to get solutions that are also
second-order accurate in time, the solution is evolved in the following
way. 

To solve Eq.(\ref{U operator}) first, the solution
of Eq.(\ref{part 2}) should be computed, which is called a system  of
Ordinary Differential Equations (ODEs). Using the two stage 
\emph{Runge-Kutta method}(Toro, 1999), which gives a second-order accurate
solution in time,

\begin{eqnarray}
K_1 = \Delta t S(t^n,U^n) \nonumber\\
K_2 = \Delta t S(t^n + \Delta t, U^n + K_1) \nonumber \\
U^{n+1} = U^n + \frac{1}{2}[K_1 + K_2].
\label{Runge-Kutta}
\end{eqnarray}

Explicit methods are much simpler to use than implicit methods. The
latter require the solution of non-linear algebraic equations at each
time step and are therefore much more expensive.
If ODEs which have no spatial variations, $\vec{F}(U) = 0$, are solved
by the implicit method, then there will be no stability
restriction on the time step. Since an explicit method is used to
solve the ODEs here, the time step is limited by the Courant condition
(Hirsch \emph{et al.}, 1992),  

\begin{eqnarray}
\Delta t = C \frac{\Delta x}{(|v^x| + C_s)_{\max}},
\label{courant condition}
\end{eqnarray}

\noindent
where $C$ is called Courant number. It can be $0 < C \leq
1$. Eq.(\ref{courant condition}) means that fastest wave at a given
time travels for at most one cell length $\Delta x$ in the sought time
step $\Delta t$. 

Second, the solution of  Eq.(\ref{Flux Part of Equation}) is computed
using a predictor-corrector time update method, as follows.

\noindent
Initially, the solution at the half time step is computed,

\begin{eqnarray}
U_{i,j,k}^{n+\frac{1}{2}} = U_{i,j,k}^{n} - \frac{\triangle t}{2
\triangle x} ((f^n)_{i+\frac{1}{2},j,k} - (f^n)_{i-\frac{1}{2},j,k} )
- \nonumber\\  
\frac{\triangle t}{2 \triangle y} ((f^n)_{i,j+\frac{1}{2},k} -
(f^n)_{i,j-\frac{1}{2},k} ) - \nonumber\\
\frac{\triangle t}{2 \triangle z} ((f^n)_{i,j,k+\frac{1}{2}} -
(f^n)_{i,j,k-\frac{1}{2}} ).
\label{half time step}
\end{eqnarray}

\noindent
Then, the solution using the full time step is computed
and new  fluxes   
$(f^{n+\frac{1}{2}})$, which are functions of $U_{i,j,k}^{n+\frac{1}{2}}$, 

\begin{eqnarray}
U_{i,j,k}^{n+1} = U_{i,j,k}^{n} - \frac{\triangle t}{\triangle x}
((f^{n+\frac{1}{2}})_{i+\frac{1}{2},j,k} -
(f^{n+\frac{1}{2}})_{i-\frac{1}{2},j,k} ) - \nonumber\\  
\frac{\triangle t}{\triangle y} ((f^{n+\frac{1}{2}})_{i,j+\frac{1}{2},k} -
(f^{n+\frac{1}{2}})_{i,j-\frac{1}{2},k} ) - \nonumber\\  
\frac{\triangle t}{\triangle z} ((f^{n+\frac{1}{2}})_{i,j,k+\frac{1}{2}} -
(f^{n+\frac{1}{2}})_{i,j,k-\frac{1}{2}} ).
\label{full time step}
\end{eqnarray}

In order to solve general relativistic hydrodynamical equation in $3-D$,
the following procedure is followed. First Eq.(\ref{source}) is
solved using second-order Runge-Kutta method, Eq.(\ref{Runge-Kutta}),
at half time step. 
Later Eq.(\ref{Flux Part of Equation}) is solved using
predictor-corrector method, Eqs.(\ref{half time step}-\ref{full time
step}), at the full time step. Finally, the source term is 
solved another half time step using  the Runge-Kutta method. At the
end of the calculation, a second-order accurate solution for
GRH equation is obtained in space and time.


\section{Adaptive-Mesh Refinement}

Adaptive mesh refinement is needed to solve some problems in 
numerical relativity, such as coalescing compact binaries, accreting compact
objects, etc. A key difficulty in calculating accurate waveforms from
these simulations lies in the fact that the length and time scales for
these problems can vary by an order of magnitude or more within the
computational domain. Adequate
resolution is needed to model  strong field regions. Adaptive
techniques are thus crucial for success in modeling such systems.

\subsection{Paramesh Package}

The \emph{Paramesh} package(MacNeice \emph{et al.}, 2000) are used   to make to
implement  AMR  and parallelization in the code.  The package runs on 
any parallel
platform which supports either the shmem(shared memory) or MPI libraries.  
This package is
a set of FORTRAN 90 routines designed to enable Adaptive Mesh
Refinement(AMR) for a parallel distributed memory machine.  

The basic AMR strategy is that the computational domain is covered
with a hierarchy of numerical sub-grids. These sub-grids  form
the nodes of a tree data-structure and  are distributed among the
processors. When more spatial 
resolution is required at some location, the highest resolution
sub-grid covering that point spawns child sub-grids (2 in 1D, 4 in 2D
and 8 in 3D), which together cover the line, area or volume of their
parent, but now with twice its spatial resolution. They refer to the
block with the highest refinement level at a given physical location
as a `leaf block'. All sub-grids have identical logical structure (\emph{i.e.},
the same number of 
grid points in each dimension, the same aspect ratios, the same number
of guard cells, etc ). They differ from each other in their physical sizes and
locations, and in their list of neighbors, parents and children. 
The package assumes that the application will use logically Cartesian
grids (\emph{i.e.}, grids that can be indexed i, j, k). It works for
1D, 2D and 3D problems.  

The package provides routines which perform all the communication
required between sub-grids and between processors. It manages the
refinement and de-refinement processes, and can balance the workload
by reordering the distribution of blocks among the processors. It
also provides routines to enforce conservation laws at the interfaces
between grid blocks at different refinement levels. Flux conservation
at different refinement level interfaces will be explained in section
\ref{Flux Conservation} .

\begin{center}
\begin{figure}
\vspace*{2.2cm}
\centerline{\epsfxsize=14cm \epsfysize=14cm
\epsffile{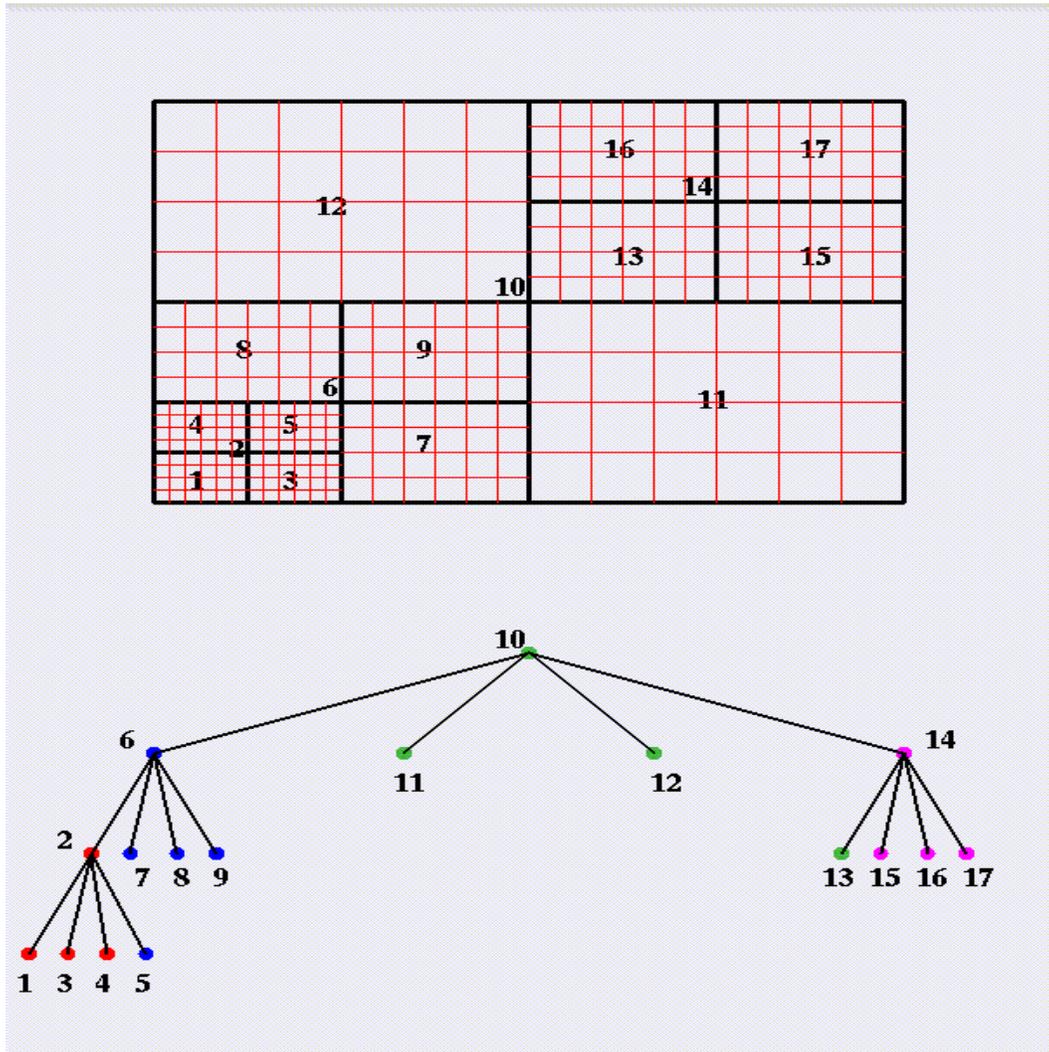}}
\caption{Representation of adaptive-mesh refinement grid  and tree
structure. Figure is taken from MacNeice \emph{et al.}, 2000.}
\label{AMR Grid sample}
\end{figure}
\end{center}

The top panel of Fig.\ref{AMR Grid sample} shows a 6 x 4 grid
structure in $2D$  on each block. The numbers assigned to each block
represent the block location in the three structure seen in the
bottom panel of Fig.\ref{AMR Grid sample}. Notice that in  Paramesh
the refinement level is not allowed to jump by more than $1$ level at any
location of the spatial domain.


\subsection{Flux Conservation}
\label{Flux Conservation}

\begin{center}
\begin{figure}
\centerline{\epsfxsize=14cm \epsfysize=14cm
\epsffile{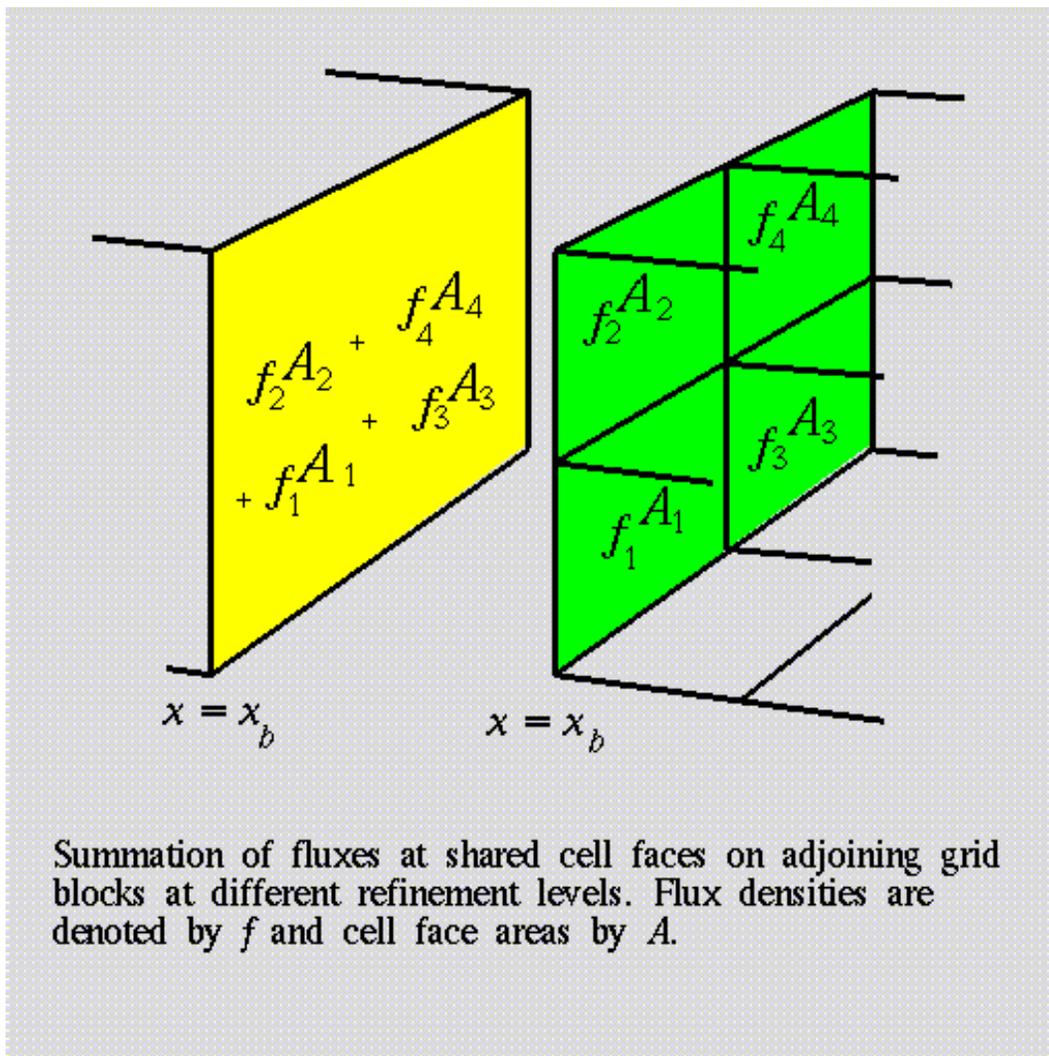}}
\caption{Flux conservation is fixed at different
refinement level boundaries. Figure is taken from MacNeice \emph{et al.}, 2000.}
\label{fixed_flux_conservation}
\end{figure}
\end{center}

Conservation of conserved variables in the hydrodynamical code is
an important physical result. To guarantee conservation of variables
on the
AMR grid, some special operation at block boundaries of different 
refinement levels are needed. Conservative hydrodynamics codes require
that  the fluxes entering or leaving a grid cell through a common
cell face shared with the $4$  refined neighbors, equal
the sum of the fluxes across the appropriate faces of the $4$ smaller
cells, Fig.\ref{fixed_flux_conservation} (MacNeice \emph{et al.}, 2000). This is
achieved by taking the  average of the fine gird fluxes and copying them
into the coarse grid 
to correct fluxes at different refinement levels. For
more information about this flux conservation, check Ref.MacNeice \emph{et al.}, 2000.


\subsection{Refinement Criteria}
\label{Refinement Criteria}

A refinement criteria needs to be defined to analyze any problem using 
Paramesh package. We have chosen to use a refinement criterion 
based on the first derivative of density, which has different  dynamics 
during the evolution for the shock tube problem. When density is refined or 
derefined, the other variables, such as pressure and velocity, will be 
refined or derefined simultaneously.  Two different 
functions are used to  define refinement criteria, which are given as 
follows:

\begin{eqnarray}
\tau_1(x,y) =  \left|\frac{h}{f} \left(\frac{\partial f}{\partial x} +
\frac{\partial f}{\partial y}\right)\right|,
\label{refinement criteria1}
\end{eqnarray}

\noindent
and

\begin{eqnarray}
\tau_2(x,y) =  \left|\frac{1}{f} \left(f(i,j) - f(i+1, j+1)\right)\right|
\label{refinement criteria2}.
\end{eqnarray}

\noindent
where Eqs.(\ref{refinement criteria1}) and (\ref{refinement
criteria2}) are the first derivative and  diagonal
derivative of a function $f$, respectively. The function $f$ can be chosen any 
variable in the simulation  depending on the change of dynamics during 
the numerical simulation.
Refinement and derefinement of blocks depend on refinement criteria
and parameters. If the value of the refinement functions, Eq.(\ref{refinement
criteria1}) or (\ref{refinement criteria2}), are bigger than refinement
parameter, {\emph ctore} specified by user based on problem, in 
any zone in the same block, that block will be refined.  If the value 
of the  refinement functions  are smaller than derefinement parameter, 
{\emph ctode}, that  block will be derefined. Refinement parameter has to 
be bigger than derefinement  parameter.
These refinement and derefinement criteria are checked on each
timestep during the numerical simulation. The  refinement criteria and 
parameters play an important role in AMR. Note that both refinement and 
derefinement in Paramesh are restricted by the requirement that the grid can 
change by only one refinement level at each block.


\section{Results From Special Relativistic Shock Tube Problem}
\label{Results From Shock Tube Problem}

The Riemann shock tube problem (Hirsch \emph{et al.}, 1992, 
Font \emph{et al.}, 2000) is used as a test case.  
In this problem, the fluid is initially in two
different thermodynamical states on either side of a membrane. The
membrane is then removed. In the test problem, it is assumed that 
the fluid initially has
$\rho_L > \rho_R$, where the subscripts $L$ and $R$ refer to the left
and right sides of the membrane. With this assumption, a rarefaction 
wave travels to
the left, and a shock wave and contact discontinuity travel to the
right. The fluid in the intermediate  state moves at a mildly relativistic 
speed which is $v = 0.725c$ to the right. Flow particles accumulate  in
a dense shell behind the shock wave compressing the fluid and  heating 
it up to values of the internal energy much larger than the rest-mass
energy. Hence, the fluid is extremely relativistic from  a thermodynamical
point of view, but only mildly relativistic dynamically. 
The Riemann shock tube
is a useful test problem because it has an exact time-dependent
solution, and tests the ability of the code to evolve both smooth and
discontinuous flows. In the case considered here, the velocities are
special relativistic and the method of finding the exact solution
differs somewhat from the standard non-relativistic shock tube.

The initial values of the fluid variables for $1D$  test are;
 $\rho_l = 10.0 $, $\rho_r = 1.0$,  $u_l = 0.0$,$u_r = 0.0$,
$p_l = 13.3$,$p_r =  0.66 . 10^{-6}$.
The computational domain 
is $0 \leq x \leq 1$  and at  $t = 0$ the membrane  is placed at $x
= 0.5$. We use the perfect fluid equation of state Eq.(\ref{flux
split21}) with $\Gamma = 5/3$.

\subsection{Uniform Solution of 3D Shock Tube}
\label{Uniform Solution from 3D Results}

The $3D$ special relativistic shock tube problem is solved using 
the initial data which is given above.  The MUSCL-left and 
right states and the minmod slope 
function with the Marquina fluxes are used. This method gives a second-order
accurate solution for the $3D$ special relativistic hydrodynamical
equations. The $L_1$ norm errors and convergence rates are computed and
our results are also compared  with the existing literature.  In
order to compare the $3D$ numerical  with $1D$ analytic
solution for the  shock tube, the initial discontinuity is set up
along the diagonal of the computational domain. The $128^3$ zones are used
with interval of length $1/\sqrt{3}$ in each directions. The diagonal of 
the cube therefore has unit length. The maximum time, $t = 0.4$, is used 
to compute analytic and numerical calculations and  
the numerical data  is extracted along the main diagonal of the cube. 

\def\baselinestretch{1.1}

\begin{table}
\caption{Comparing our code $L_1$ norm errors with the other hydrodynamic code
results from $3D$.}
$$ \vbox{ \offinterlineskip \vskip5pt
  \def\qq{\hskip0.5em}
  \def\laststrut{\vrule depth6pt width0pt}
  \def\titlestrut{\vrule height12pt depth6pt width0pt}
\halign {#\vrule\strut &\quad#\hfil
       &&\qq#\vrule &\qq\hfil#\hfil \cr
\noalign{\hrule}
 & \multispan{9}\hfil L1 norm error for shock tube problem \hfil\titlestrut 
  &\cr \noalign{\hrule}  
& Code && $npts$ && $L_1(\rho)$&& $L_1(p)$&& $L_1(v)$  &\cr
  \noalign{\hrule}

& OUR   && $32^3$ && 0.188 && 0.1798 && 1.82E-2 &\cr
& CODE && $64^3$ && 9.813E-2&& 9.199E-2 && 6.178E-3  &\cr
&     && $128^3$ && 5.439E-2 && 4.595E-2 && 3.86E-3 &\cr
  \noalign{\hrule}
  \noalign{\hrule}
& Font,Miller && $128^3$ && 9.23E-2 && 7.98E-2 && 9.66E-3 &\cr
& Suen,Tobias &&  &&  &&  &&  &\cr
  \noalign{\hrule}
  \noalign{\hrule}
&    && $40^3$ && 1.1E-1 && 8.0E-2 && 0.9E-2 &\cr
&    && $60^3$ && 9.8E-2 && 5.2E-2 && 1.1E-2 &\cr
& GENESIS && $80^3$ && 9.2E-2 && 4.5E-2 && 1.1E-2 &\cr
&  Aloy \emph{et al.}, 2000  && $100^3$ && 7.0E-2 && 3.7E-2 && 7.0E-3 &\cr
&    && $150^3$ && 4.8E-2 && 2.5E-2 && 5.0E-3 \laststrut &\cr
\noalign{\hrule\vskip4pt} 
  \noalign{\vskip5pt} }}$$
\label{MUSCL-shock1 3D}
\end{table}

\begin{table}
\caption{Convergence rate  from $3D$ special relativistic hydro code.}
$$ \vbox{ \offinterlineskip \vskip5pt
  \def\qq{\hskip0.6em}
  \def\laststrut{\vrule depth6pt width0pt}
  \def\titlestrut{\vrule height12pt depth6pt width0pt}
\halign {#\vrule\strut &\quad#\hfil
       &&\qq#\vrule &\qq\hfil#\hfil \cr
\noalign{\hrule}
 & \multispan{7}\hfil Convergence rate for the shock tube problem \hfil\titlestrut 
  &\cr \noalign{\hrule}  
& $npts - 2 * npts$ && $r(\rho)$&& $r(p)$&& $r(v)$  &\cr
  \noalign{\hrule}
& $32^3-64^3$ &&  0.94 && 0.9668 && 1.563   &\cr
& $64^3-128^3$ && 0.851 && 1.001  && 0.68  \laststrut &\cr
\noalign{\hrule\vskip4pt} 
  \noalign{\vskip5pt} }}$$   
\label{MUSCL-shock2 3D}
\end{table}

\def\baselinestretch{1.75}

Table \ref{MUSCL-shock1 3D} shows the $L_1$-norm errors of different
hydrodynamical quantities. These are the density, pressure and
velocity. We present our results, those from Font \emph{et al.}, 2000, and
GENESIS code(Aloy \emph{et al.}, 2000) in Table \ref{MUSCL-shock1 3D}; 
all of these  are $3D$ simulations. Our code and the code of 
Ref.Font \emph{et al.}, 2000 use the Marquina method with MUSCL
left and right states which give a second-order accurate solution. The
Genesis code uses Marquina method with the PPM reconstruction
procedure which gives third-order accurate solutions. 
Table \ref{MUSCL-shock1 3D} shows that the results are
comparable. Another difference between the Genesis and our code is that we
compute $L_1$-norm errors at $t=0.4$ but they compute at $t=0.5$. Marquina
method with PPM left and right states is computationally more 
expensive than our code. Because it has more reconstruction
process and it needs to more evolution step to get the third-order
accuracy in time. The PPM interpolation algorithm described in
Colella \emph{et al.}, 1984 gives monotonic conservative parabolic profiles of
variables within a numerical zones. In the relativistic version of
PPM, the original interpolation algorithm is applied to zone averaged
values of the primitive variables $(\rho, v, p)$, which are obtained
from zone averaged values of the conserved variables $U$. 

\begin{center}
\begin{figure}
\centerline{\epsfxsize=14cm \epsfysize=14cm
\epsffile{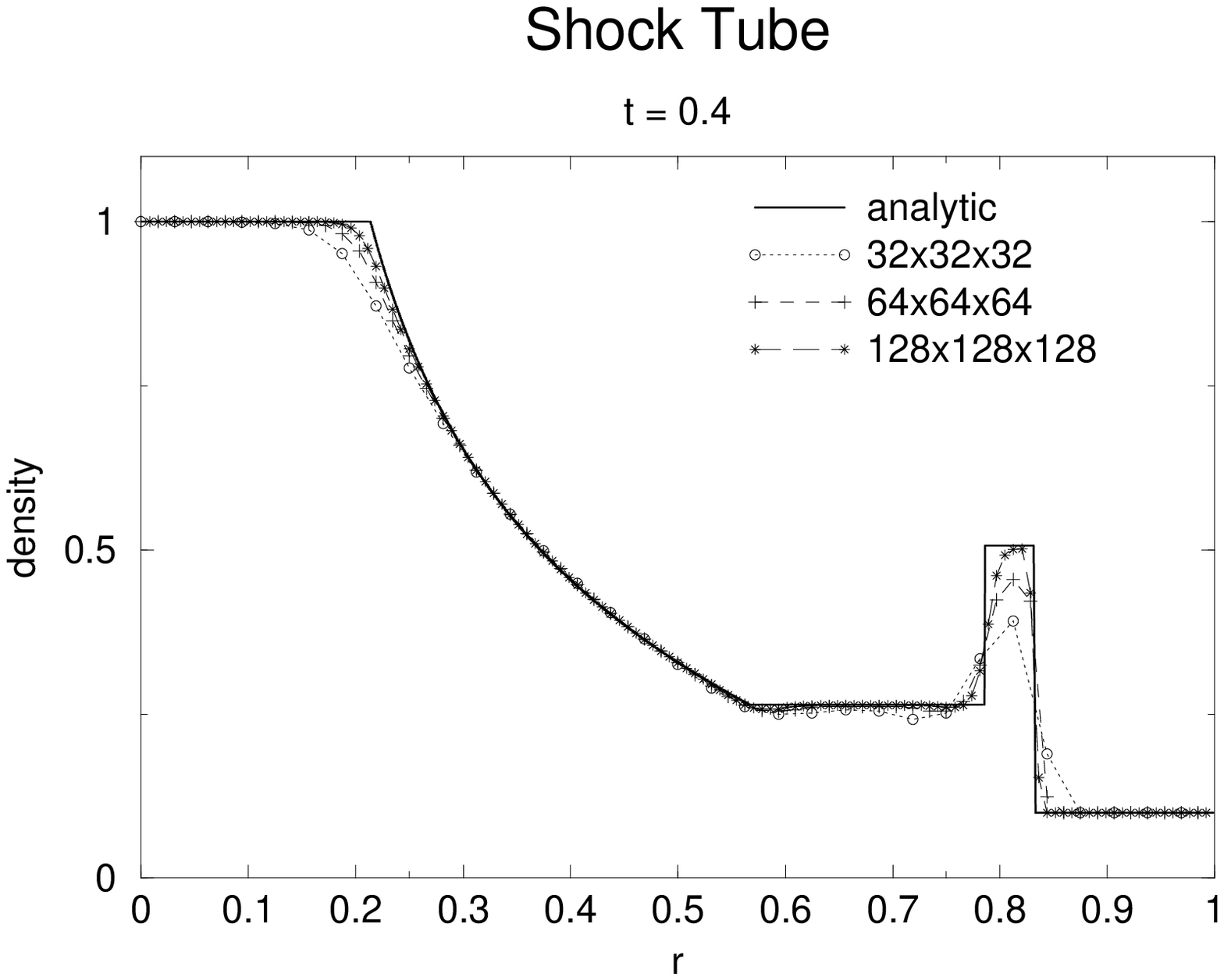}}
\caption{The analytic and numerical solutions of the relativistic
shock tube problem for $3D$. }
\label{MUSCL-shock3 3D}
\end{figure}
\end{center}

The convergence factors are shown in Table \ref{MUSCL-shock2 3D}.
It should approach $1$ when the higher order methods are used.  
Fig. \ref{MUSCL-shock3 3D} shows the result
obtained for different numbers of points with this method. The solid 
line represents the exact solution while different
symbol represents the numerical results for different resolution for
the density. As can be seen from Fig. \ref{MUSCL-shock3 3D} that when the number 
of resolutions are increased, numerical solution converge to analytic one.

\def\baselinestretch{1.2} 


\subsection{AMR Results}

\begin{center}
\begin{figure}  
   \epsfxsize=14.0cm \epsfysize=14.0cm
     \epsfbox{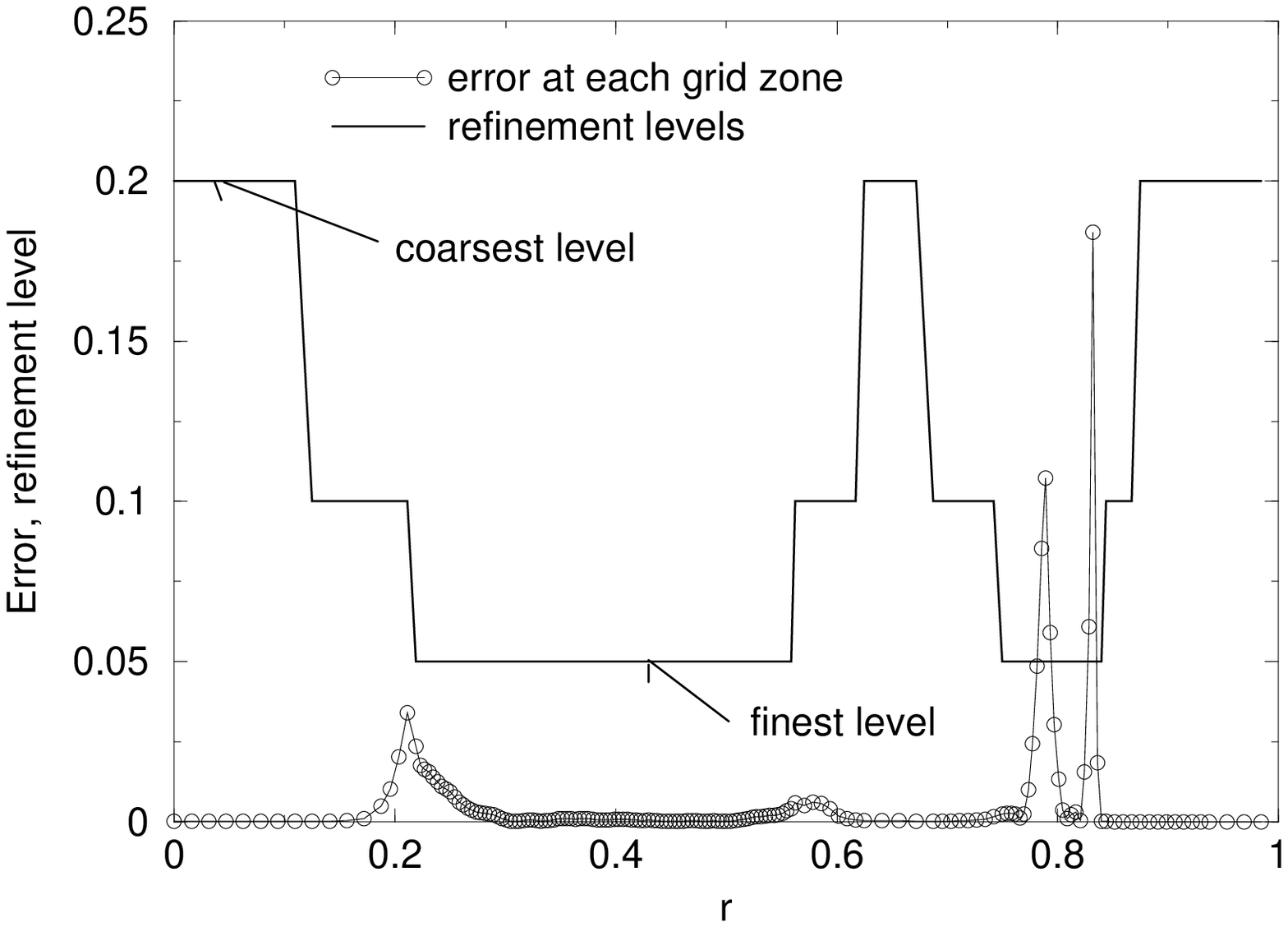}
\caption{Plotting the error and the refinement levels in each zone
     Eq.(\ref{refinement criteria1}) is used as the  refinement
     function with parameters: ctore(refinement parameter) =
     $0.035$, ctode(derefinement parameter) = $0.005$} 
\label{AMR run 1}
\end{figure}
\end{center}

We have first carried out  simulations of the $2D$ shock tube using 3 levels 
of refinements. To do this problem, initial discontinuity has been  placed 
along the diagonal axis of the computational domain (a square). The 
evolution of the system along the direction orthogonal to this diagonal 
is purely one dimensional problem. It allows us to compare one-dimensional 
analytic solution  with the numerical results. The results of shock 
tube problem  
with AMR are compared with analytic solution at each grid zone of the 
computational domain.
The error in each grid zone and the refinement levels are plotted
in Fig.\ref{AMR run 1}. The finest level is replaced around $r = 0.8$ where
shock wave is produced. Eq.\ref{refinement
criteria1} is used as a  refinement function. Refinement parameters 
for this problem are: ctore(refinement parameter) = $0.035$,
ctode(derefinement parameter) = $0.005$. 

\begin{center}
\begin{figure}  
     \epsfxsize=14.0cm \epsfysize=14.0cm
     \epsfbox{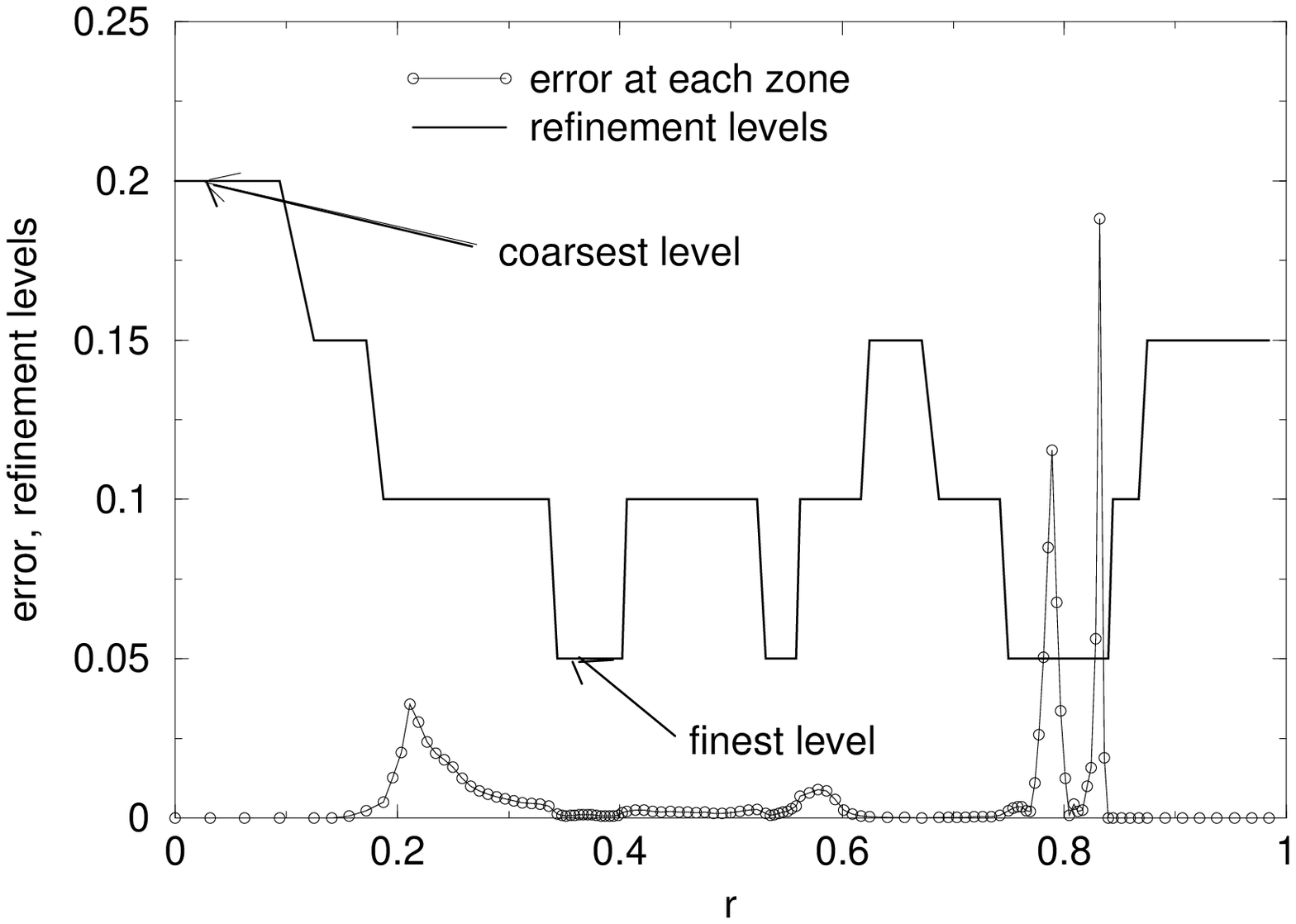}
    \caption{ Plotting the error and refinement levels in each zone.
     Eq.(\ref{refinement criteria2}) is used as the refinement
     function with  parameters: ctore(refinement parameter) =
     $0.035$, ctode(derefinement parameter) = $0.005$}
\label{AMR run 3}
\end{figure}
\end{center}

The refinement function in Eq.\ref{refinement criteria2} is used
to solve the same problem  with the same refinement parameters 
of Fig.\ref{AMR run 1}. The 
numerical results are given at Fig.\ref{AMR run 3}. The results show that 
Fig.\ref{AMR run 3} is more derefined regions and use less number of 
resolutions.


In this part of test problem same refinement function is used in 
Fig.\ref{AMR run 3}  but refinement parameters 
for this problem are: ctore(refinement parameter) = $0.035$,
ctode(derefinement parameter) = $0.007$.

\begin{center}
\begin{figure}  
     \epsfxsize=14.0cm \epsfysize=14.0cm
     \epsfbox{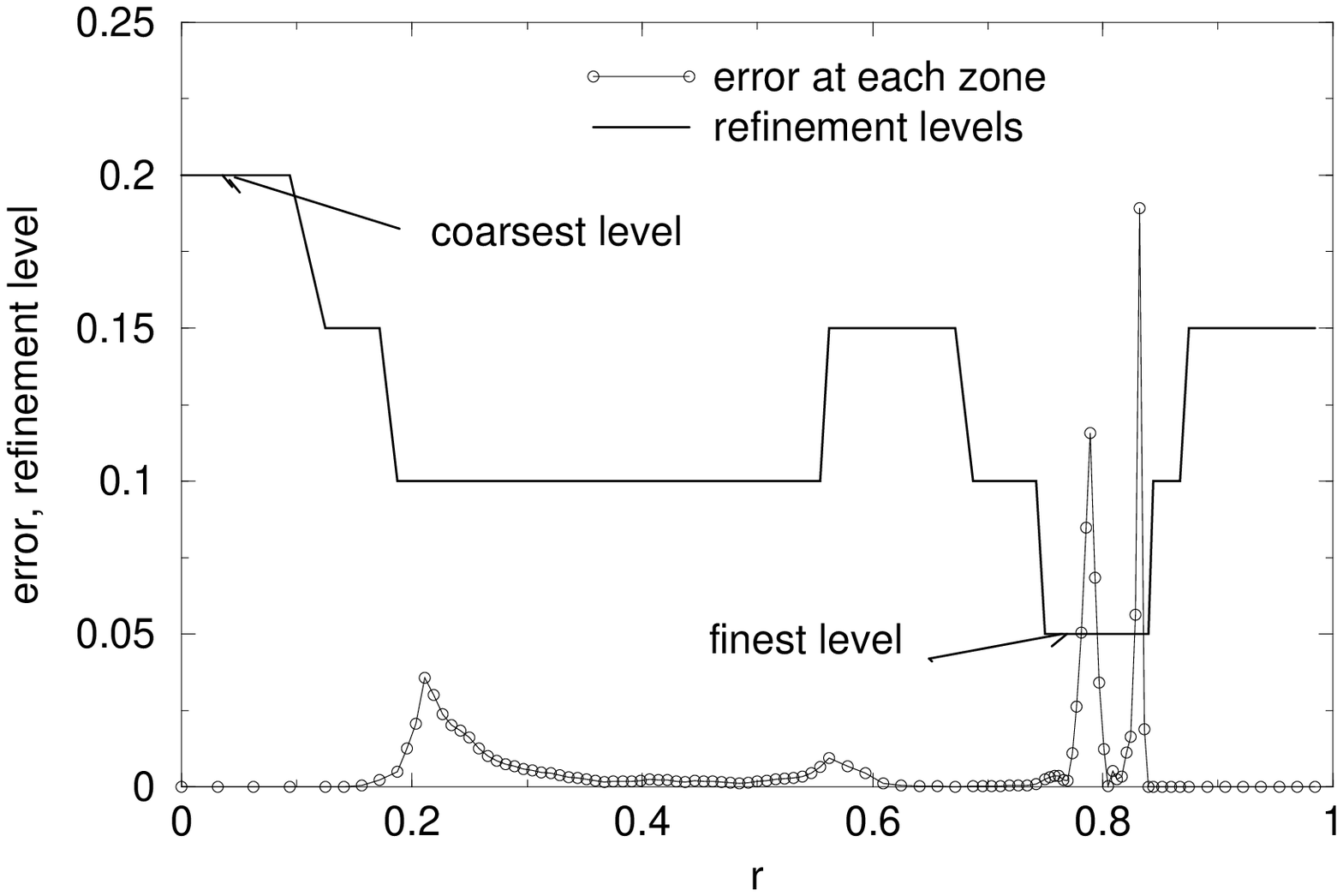} 
     \caption{Plotting the error and refinement levels in each zone.
     Eq.(\ref{refinement criteria2}) is used as the refinement
     function with parameters: ctore(refinement parameter) =
     $0.035$, ctode(derefinement parameter) = $0.007$}
\label{AMR run 5}
\end{figure}
\end{center}

\noindent
It is seen from Figs.\ref{AMR run 1}, \ref{AMR run 3} and \ref{AMR run 5} that 
refinement levels are different for each cases. Fig.\ref{AMR run 5} 
has more derefine  regions
than  the others. If the more derefine regions are observed, this indicates
that less number of points is used during the evolutions.
The highest refinement level is replaced at discontinuity. 
As a conclusion, AMR grid structure depends on refinement 
functions and parameters which need to be defined carefully for 
different problems.

\begin{center}
\begin{figure}  
     \epsfxsize=14.0cm \epsfysize=14.0cm
     \epsfbox{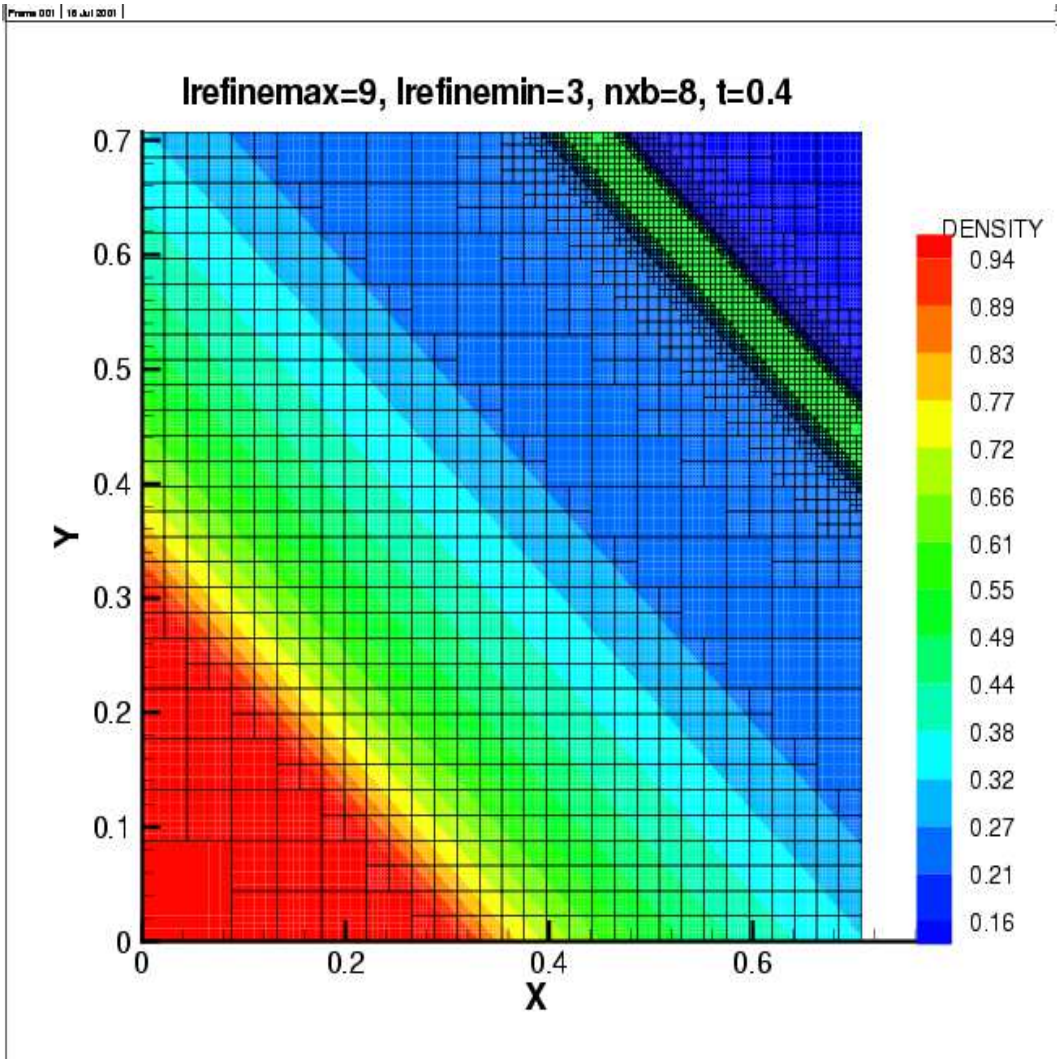}
\caption{Density  is plotted in the x-y plane. Each square
     represents the block structure and each block has 8 points in a each
     direction. Eq.(\ref{refinement criteria2}) is used  as the 
     refinement function with refinement parameters:
     ctore(refinement parameter) = $0.035$, ctode(derefinement
     parameter) = $0.007$.}  
\label{AMR run 7}
\end{figure}
\end{center}

\begin{center}
\begin{figure}  
     \epsfxsize=14.0cm \epsfysize=14.cm
     \epsfbox{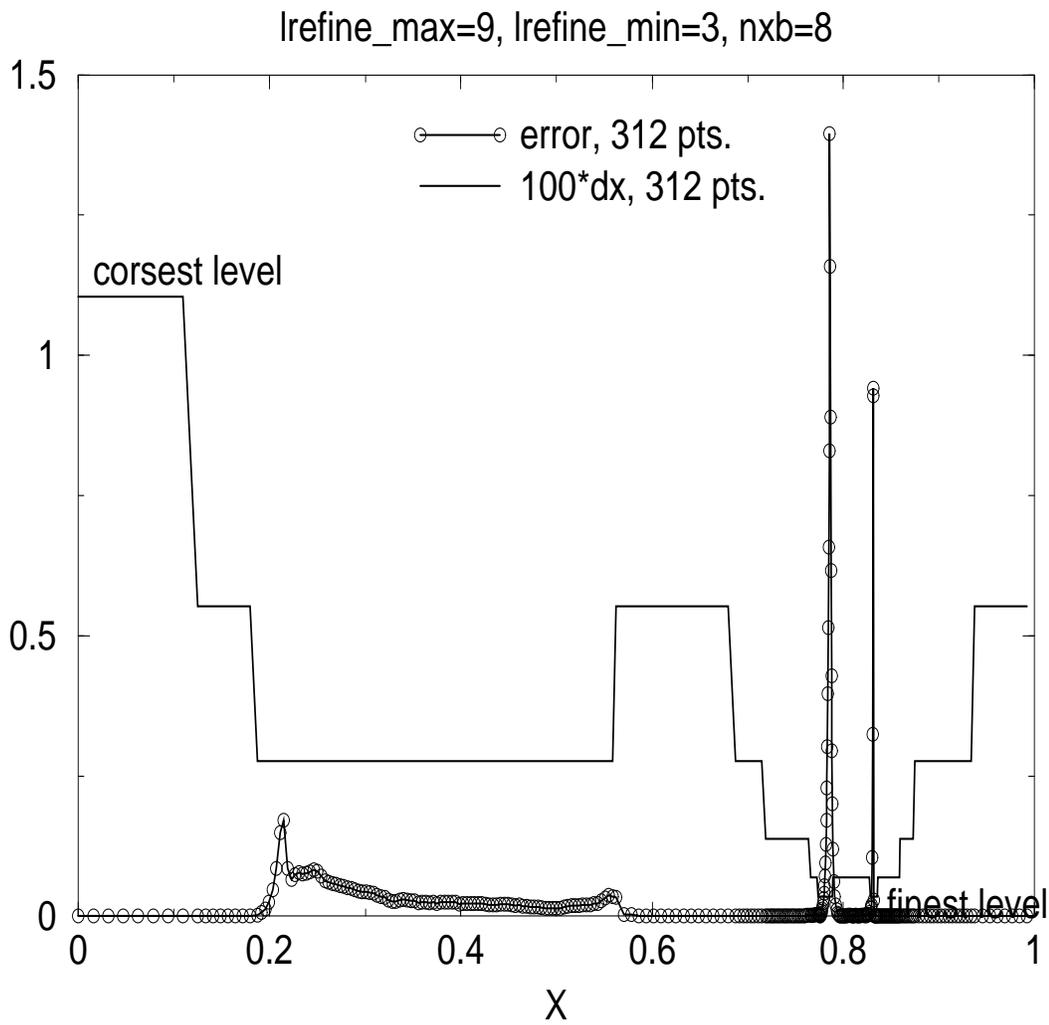}
\caption{We plot the  error
     and refinement levels in each zone to see
     the behavior of the refinement levels with errors. Eq.(\ref{refinement
     criteria2}) is used  as the refinement function with refinement
     parameters: ctore(refinement parameter) = $0.035$,
     ctode(derefinement parameter) = $0.007$.}
\label{AMR run 8}
\end{figure}
\end{center}

Here, the Riemann shock tube problem is solved in $2D$ using seven different 
refinement levels with the best refinement parameters which 
are ctore(refinement parameter) = $0.035$ and 
ctode(derefinement parameter) = $0.007$. In Fig.\ref{AMR run 7}, 
density is displayed  on the $x-y$ plane at the time when the shock 
tube structure is well-developed. The AMR
run of the shock tube problem creates a nice grid structure 
that follows the density gradients in the problem well.
In Fig.\ref{AMR run 8}, the  error function for
each grid zone is given  with refinement level to see the behavior of
the adaptive grid during the evolution. We have also looked at the movie  
for this problem to see the behavior of refine ad derefine grid during the 
calculations. The finest grid follows the shock wave and some derefine regions 
are produced at contact discontinuity which is a constant state.



\begin{figure}  
     \centerline{\epsfxsize=14.cm \epsfysize=14.cm
     \epsfbox{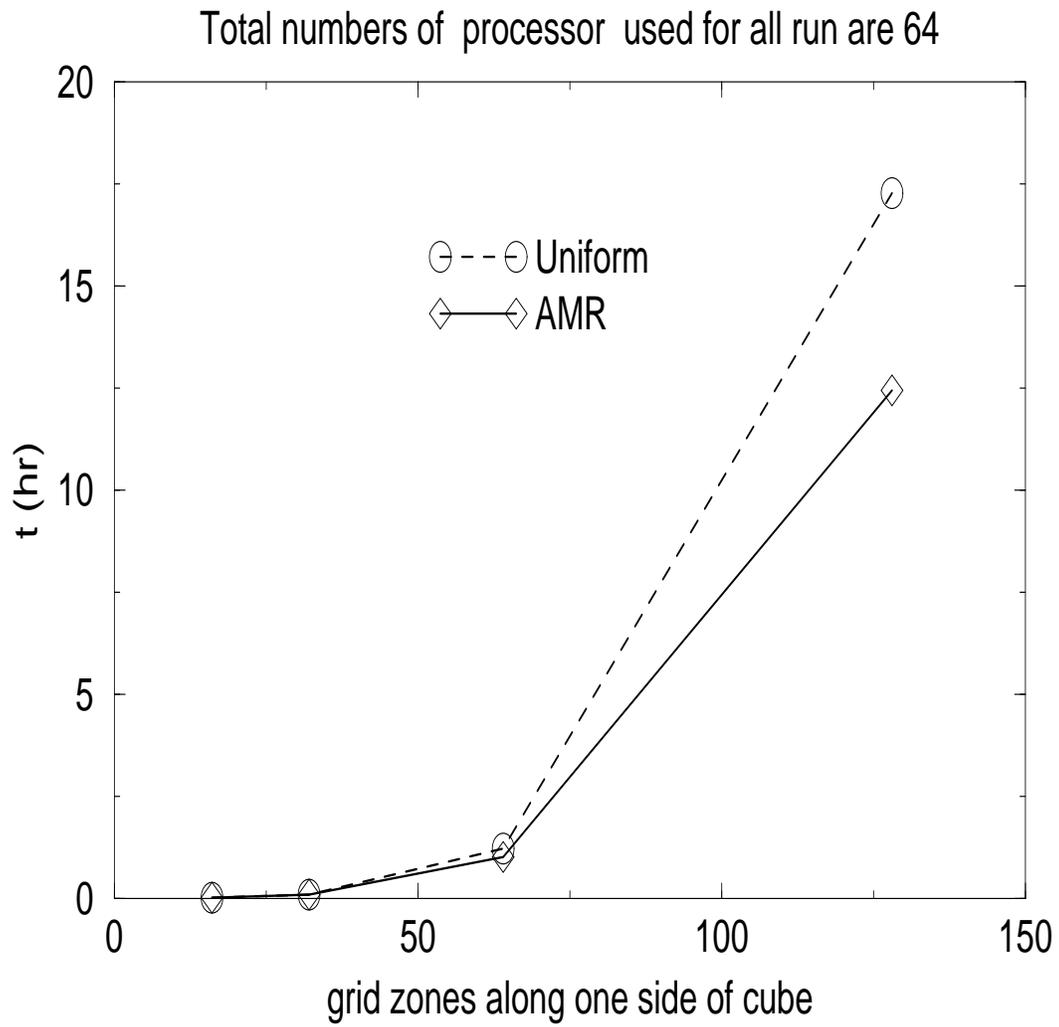}
}
\caption{The computing time for AMR and uniform grid runs of the $3D$
     shock tube problem. The red line is the time
     per processor for AMR and the black line is the time per
     processor for uniform grid runs.}
\label{AMR run 9}
\end{figure}

It is interesting to compare the computing time for AMR and uniform runs
 in $3D$. Fig.\ref{AMR run 9} illustrates the  time  per
processor  using  different resolution for
AMR and uniform runs. The times for the uniform and AMR cases  are the same
for low resolutions, $16^3$ and $32^2$, 
because  the whole computational domain is refined to
 the maximum level at initial time. It means that there is no
 AMR in these cases. But the computing
 times for the resolutions $64^3$ and $128^3$
in AMR are less than for uniform grids. The computing time for lower
resolution is affected by overhead in the AMR routines. Saving the 
computational cost in this test problem for $128^3$  is approximately $30\%$.
Increasing  the size of computational domain, the number of refinement levels
and the number of resolutions increase saving of the computational cost in the 
shock tube problem. Increasing these futures allow us to see more lower levels 
grids and more derefine region when shock wave moves to left and rarefaction 
wave move to right.

\begin{figure}  
     \centerline{\epsfxsize=10.cm \epsfysize=10.cm
     \epsfbox{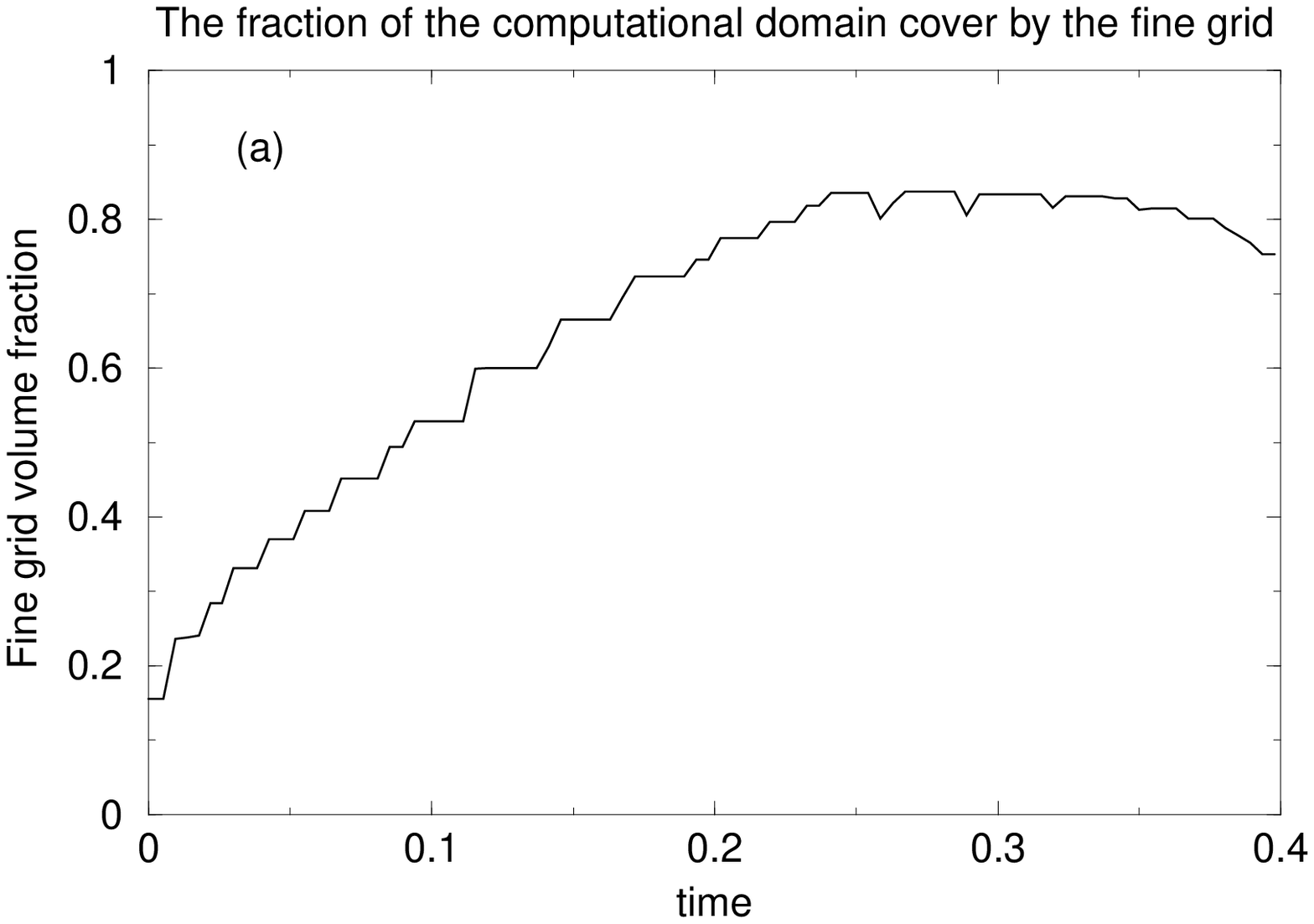}
}
     \centerline{\epsfxsize=10.cm \epsfysize=10.cm
     \epsfbox{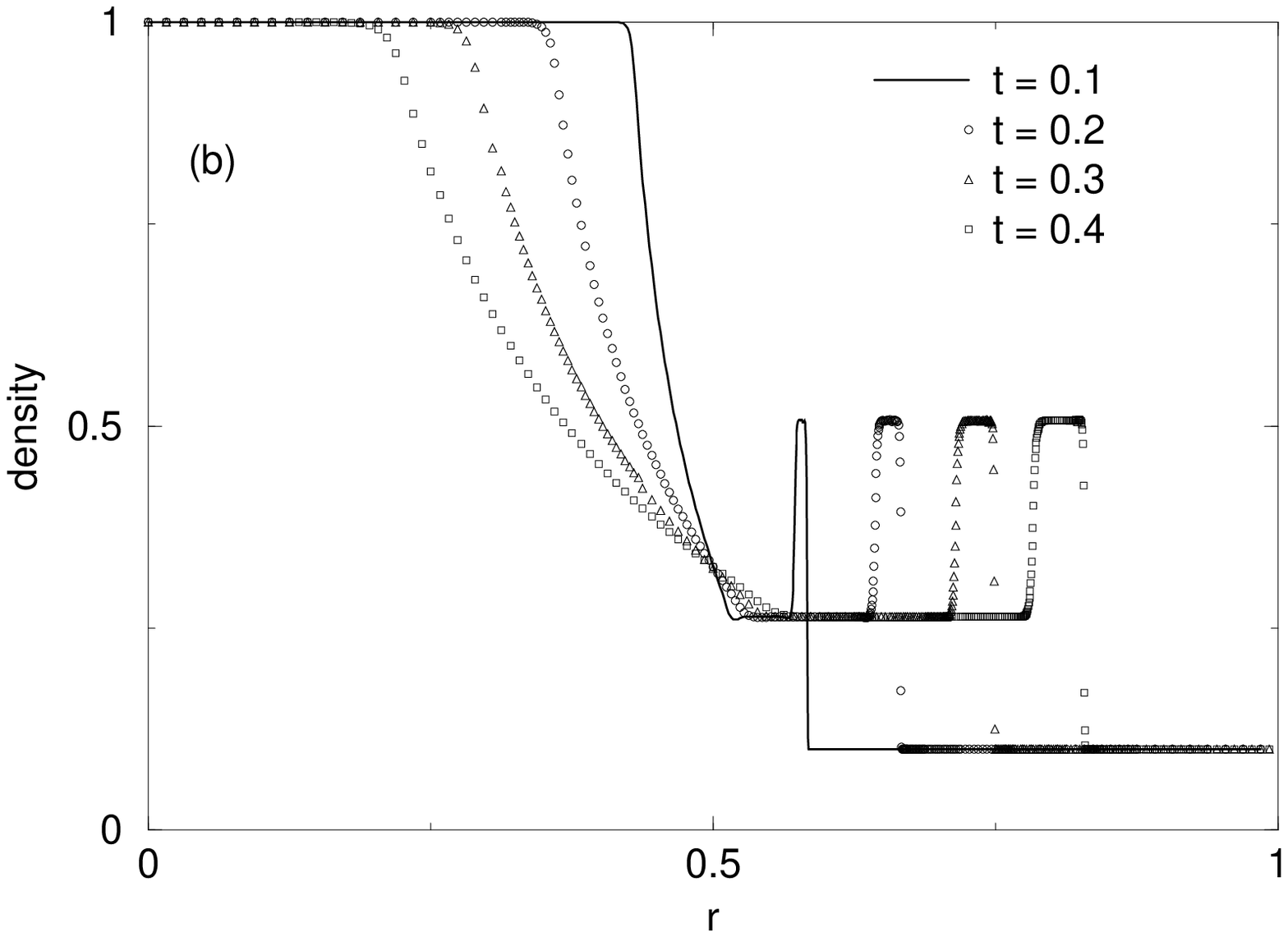}
}	
 
\caption{(a): The fraction of the volume covered by the fines
     grid for AMR run of the $3D$ shock tube problem. (b): 
     the density of the shock tube problem at different times.}
\label{AMR run 10}
\end{figure}

Fig. \ref{AMR run 10} depicts the fine grid volume fraction in the AMR
simulation with $128^3$ zones. Initially, less than 15\% of the computational
volume was covered by the fine grid. The fine grid volume fraction increases
as the shock wave 
propagates to the left and the rarefaction wave propagates to the right. 
Although  derefinement in the  computational domain  occurs around the 
origin at  $t=0.25$  the total number of grid cells continues to increase 
as the shock and rarefaction waves move to the right and left, respectively, 
as shown in Fig. \ref{AMR run 10}. The computational volume is more derefined
than refined around  $t=0.3$. The numbers of refined and
derefined zones depend on the refinement criteria and parameters
used in numerical simulations. 

In these kinds of test simulations, the saving in computational cost of 
the AMR run over a unigrid run are modest since the wavelength of the 
wave, $\lambda$  is 
comparable to the dimension $L$  of the computational domain.
In the simulation of the real astrophysical problem, it is
expected that wavelength is much smaller than the dimension of computational
domain, resulting in significant computational savings. For a full simulation 
of binary coalescence or accretion disk around the compact objects, one will 
need to handel multiple scales to evolve dynamics of the sources and 
propagation of the waves that develop. While the sources are expected 
to require high resolution, particularly near regions of shocks or 
shear layers, the waves can likely be handled with lower resolution.


%


\section{General Relativistic Hydrodynamical Test Problem}
\label{General Relativistic Hydrodynamical Test Problem}

In this section, we first introduce the Schwarzschild geometry in
spherical coordinates to define sources in the general relativistic
hydrodynamical equations. Then the accretion disk
problems are solved , which have been analytically analyzed, 
to test the full GRH code in $2D$ on the equatorial plane.

\subsection{Schwarzschild Black Hole}
\label{Schwarzschild Black Hole}

The Schwarzschild solution determined by the mass $M$ gives the geometry 
in outside of a spherical  star or black hole.
The Schwarzschild spacetime metric in spherical coordinates is

\begin{eqnarray}
ds^2 = -(1 - \frac{2 M}{r}) dt^2 + (1 - \frac{2 M}{r})^{-1} dr^2 + \nonumber \\
r^2 d\theta^2 + r^2 sin^2 \theta d\phi^2 .
\label{line element polor coordinate}
\end{eqnarray}

\noindent
It behaves badly near $r=2M$; there the first term becomes zero and
the second term becomes infinite in Eq.(\ref{line element polor
coordinate}). That radius $r=2M$ is called the Schwarzschild radius  or
the Schwarzschild horizon. 

\noindent
The spacetime metric for this line element is as follows:

\begin{eqnarray}
g_{\mu \nu} = \left(
\begin{array}{cccc}
-(1 - \frac{2 M}{r}) & 0 & 0 & 0 \\
0 & (1 - \frac{2 M}{r})^{-1} & 0 & 0 \\
0 & 0 & r^2 & 0 \\
0 & 0 & 0 & r^2 sin^2 \theta
\end{array}
\right) .
\label{g metric}
\end{eqnarray}

\noindent
The lapse function and shift vector for this metric is given bellow:

\begin{eqnarray}
\beta^r = 0.0, \;\;\;  \beta^{\theta} = 0.0, \;\;\;\; \beta^{\phi} =
0.0 ,
\nonumber\\
\alpha = (1 - \frac{2 M}{r})^{1/2} .
\label{lapse and shift}
\end{eqnarray}


\subsection{The Source Terms For Schwarzschild Coordinates}
\label{Source Termes For Schwarzschild Coordinates}

The gravitational sources for the GRH equations are given by
Eq.(\ref{matrix form of source 
vector}). In order to compute the sources in Schwarzschild coordinates for
different conserved variables, Eq.(\ref{matrix form of source vector}) 
can be rewritten as,

\begin{eqnarray}
\vec{S} =  \left( 
\begin{array}{c}
0 \\
\alpha \sqrt{\gamma} T^{\mu \nu} g_{\nu \sigma} \Gamma^{\sigma}_{\mu
r}  \\
\alpha \sqrt{\gamma} T^{\mu \nu} g_{\nu \sigma} \Gamma^{\sigma}_{\mu
\theta}  \\
\alpha \sqrt{\gamma} T^{\mu \nu} g_{\nu \sigma} \Gamma^{\sigma}_{\mu
\phi}  \\
\alpha \sqrt{\gamma} (T^{\mu 0} \partial_{\mu} \alpha - \alpha T^{\mu
\nu} \Gamma^{0}_{\mu \nu}) 
\end{array} 
\right). 
\label{matrix form of source vector2}
\end{eqnarray}

\noindent
It is seen in Eq.(\ref{matrix form of source vector}) that the source for
conserved density, $D$, is zero but the other sources depend on
the components of the stress energy tensor,  Christoffel symbols, and
$4$-metric.  
After doing some straightforward calculations, the sources can be
rewritten in Schwarzschild coordinates
for each conserved variable with the following form:

\noindent
The source for the momentum equation in the radial direction is

\begin{eqnarray}
\alpha \sqrt{\gamma} T^{\mu \nu} g_{\nu \sigma} \Gamma^{\sigma}_{\mu
r} = \frac{1}{2} \alpha \sqrt{\gamma} (T^{tt} \partial_{r} g_{tt} + T^{rr} \partial_{r} g_{rr} + \nonumber \\
T^{\theta \theta} \partial_{r} g_{\theta \theta}  + T^{\phi \phi}
\partial_r g_{\phi \phi}).
\label{explicit source in radial direction}
\end{eqnarray}

\noindent
The source for the momentum equation in the $\theta$ direction is

\begin{eqnarray}
\alpha \sqrt{\gamma} T^{\mu \nu} g_{\nu \sigma} \Gamma^{\sigma}_{\mu
\theta} = \frac{1}{2} \alpha \sqrt{\gamma} T^{\phi \phi}
\partial_{\theta} g_{\phi \phi}.
\label{explicit source in theta direction}
\end{eqnarray}

\noindent
The source for the momentum equation in the $\phi$ direction is

\begin{eqnarray}
\alpha \sqrt{\gamma} T^{\mu \nu} g_{\nu \sigma} \Gamma^{\sigma}_{\mu
\phi} = 0.0.
\label{explicit source in phi direction}
\end{eqnarray}

\noindent
The source for the energy equation is

\begin{eqnarray}
\alpha \sqrt{\gamma} (T^{\mu 0} \partial_{\mu} \alpha - \alpha T^{\mu
\nu} \Gamma^{0}_{\mu \nu}) = \nonumber \\ 
\alpha \sqrt{\gamma} (T^{r t}
\partial_{r} \alpha - \alpha T^{r t} g^{tt} \partial_{r} g_{tt}).
\label{explicit source for energy equation}
\end{eqnarray}

The non-zero components of the stress-energy tensor in Schwarzschild
coordinates can be computed by Eq.(\ref{des 7});
they are 

 \begin{eqnarray}
T^{tt} = \rho h \frac{W^2}{\alpha^2} + P g^{tt} \nonumber \\
T^{rr} = \rho h W^2 (v^r)^2 + P g^{rr} \nonumber \\
T^{\theta \theta} = \rho h W^2 (v^{\theta})^2 + P g^{\theta \theta} \\ 
T^{\phi \phi} = \rho h W^2 (v^{\phi})^2 + P g^{\phi \phi} \nonumber \\ 
T^{tr} = \rho h \frac{W^2}{\alpha} v^r . \nonumber
\label{component of stres energy tensor}
\end{eqnarray}


\subsection{Geodesics Flows}
\label{Geodesics Flows}

As a general relativistic test problem, an accretion of dust 
particles onto a black
hole is solved. The exact solution for pressureless dust  is 
given in Ref.Hawley \emph{et.al}, 1984. 
This problem is evolved in $2D$ in spherical
coordinates at constant $\theta = \pi/2$, which is  the equatorial
plane. Accordingly, the spatial numerical domain is the
$(r, \phi)$ plane. The $2.4 \leq r \leq 20$ and  $0 \leq \phi
\leq 2\pi$ are  used for the computational domain. The initial conditions for
all variables  are chosen to have negligible values except the outer
boundary $(r = 20 M)$ where gas is continuously injected radially 
with  the  analytic density and velocity. They are formulated as:

\begin{eqnarray}
v^r = \sqrt{\frac{2M}{r}} \sqrt{1 - \frac{2M}{r}},
\label{geodesic8}
\end{eqnarray}

\noindent
and

\begin{eqnarray}
\rho = \frac{1}{W} \frac{d}{r^2 (\frac{2 M}{r})^{\frac{1}{2}} (1 - \frac{2
M}{r})^{\frac{1}{2}}},
\label{density for free fall}
\end{eqnarray}

\noindent
where $W$ is the Lorentz factor and is given by

\begin{eqnarray}
W = \frac{1}{(1 - \frac{2 M}{r})^{\frac{1}{2}}}.
\label{lorentz factor for free fall}
\end{eqnarray}

\noindent
here $M$ is mass of black hole

Throughout the calculation,
whenever values at the outer boundary are needed, the analytic
values are used. The code is run until  a steady state solution is reached 
using outflow boundary conditions at $r = 2.4$, inflow boundary
conditions at $r = 20$ and the periodic boundary in the  $\phi$
direction are used. No dependence on $\phi$ is found throughout the entire 
simulation.

\begin{center}
\begin{figure}
\centerline{\epsfxsize=14.cm \epsfysize=14.cm
\epsffile{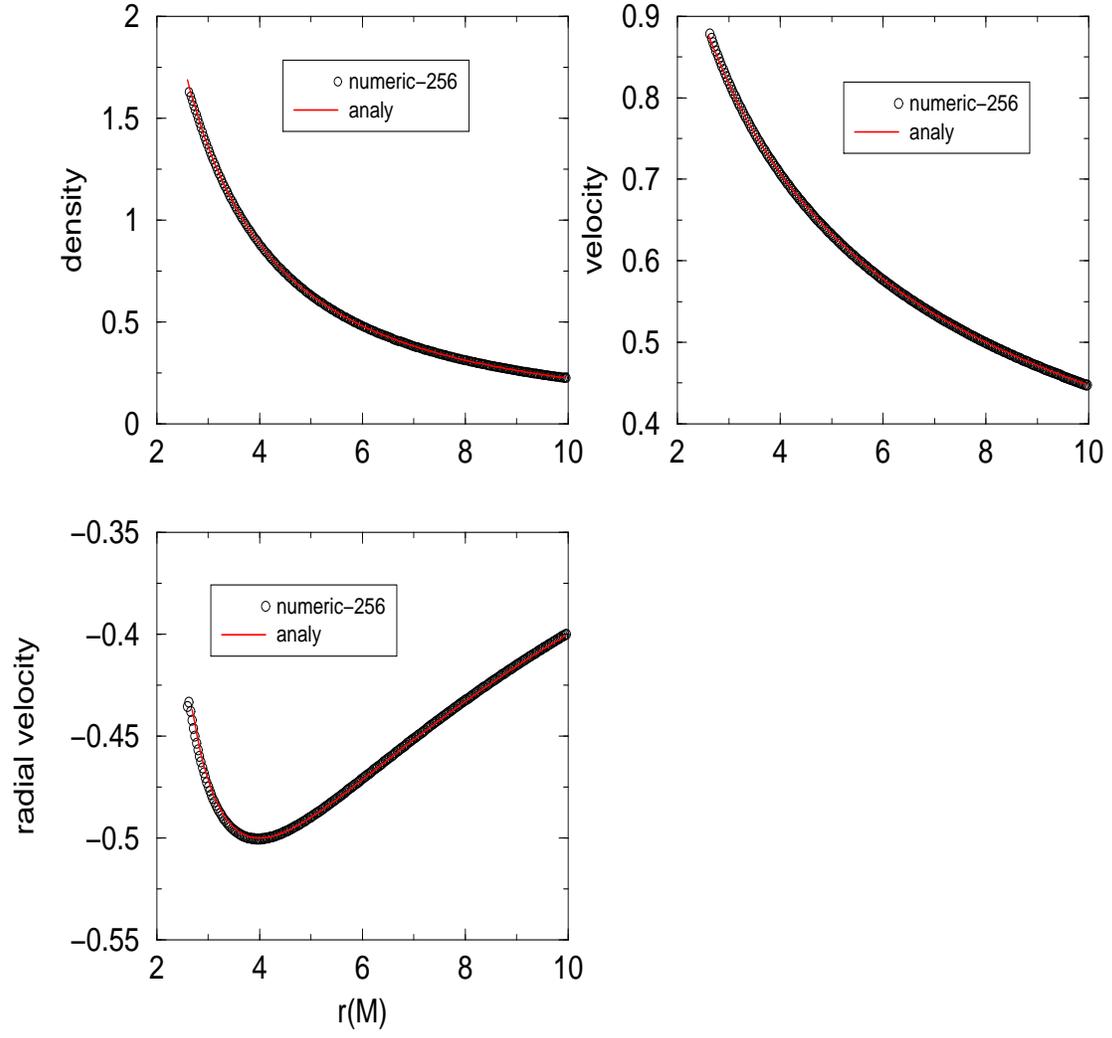}}
\caption{The analytic solutions (red solid lines)
with the numerical solutions (black circles) using $256$ zones in the
radial direction for thermodynamical variables, $\rho$, $v$, and
$v^r$.  }
\label{free falling plot for all variables}
\end{figure}
\end{center}

Fig. \ref{free falling plot for all variables} depicts
the rest-mass density $\rho$, absolute velocity $v = (v^i v_i)^{1/2}$
and radial velocity $v^r$ 
as a function of radial coordinate at a fixed angular
position. The numerical solution agrees well  with the analytic
solution.  Some convergence tests with the SR Hydro code 
are conducted. They confirmed that it is
second order convergent. The convergence test are also made on the 
this problem 
to test the behavior of the source terms in the GR Hydro code.
The analytic values of the accretion problem are compared with numerical 
results. The convergence results are given at 
Table\ref{convergence data for free fall}. 
It is noticed that code gives roughly second order convergence.

The conservation form of the general relativistic hydrodynamical equations 
are solved. From \ref{desired equation}, the general form of a conservation 
law is expressed by stating that the variation per unit time of this conserved 
equation within the volume is zero.
Then it is expected that conserved variables must be conserved to machine 
accuracy, $ \sim 10^{-16}$, at each time step. The  results 
of conservation variables in the numerical test problem show 
that these variables are conserved to 
machine accuracy.

\def\baselinestretch{1.2}

\begin{center}
\begin{table}
\caption{$L_1$ norm error and convergence factors are given  for
different resolutions.}
$$ \vbox{ \offinterlineskip \vskip8pt
  \def\qq{\hskip0.15em}
  \def\laststrut{\vrule depth7pt width0pt}
  \def\titlestrut{\vrule height12pt depth6pt width0pt}
\halign {#\vrule\strut &\quad#\hfil
       &&\qq#\vrule &\qq\hfil#\hfil \cr
\noalign{\hrule}
 & \multispan{7}\hfil Convergence Test \hfil\titlestrut 
  &\cr \noalign{\hrule}  
& \# of points &&  \# of time step && $L_1$ norm error && Convergence
  factor &\cr  
  \noalign{\hrule}
& $32$ && $1$ && $1.675E-5$ &&   \laststrut &\cr
& $64$ && $2$ && $4.0811E-6$ && 4.105 \laststrut &\cr
& $128$ && $4$ && $9.6456E-7$ && 4.23 \laststrut &\cr
& $256$ && $8$ && $2.1372E-7$ && 4.51 \laststrut &\cr
\noalign{\hrule\vskip4pt} 
  \noalign{\vskip5pt} }} $$
\label{convergence data for free fall}
\end{table} 
\end{center}

\subsection{Circular Motion of  Test Particles}
\label{Circular Motion of The Test Particle}

The circular motion of a fluid on the equatorial plane will be simulated.
To do this, the circular flow is set up with negligible pressure in the 
equatorial plane, in which angular velocity at each radial direction
$r$ is given by:

\begin{eqnarray}
v^{\phi} =\frac{1}{\sqrt{(1 - \frac{2M}{r})}} \sqrt{\frac{M}{r^3}}.
\label{equation for circular motion10}
\end{eqnarray}

\noindent
It is called a Keplerian velocity. 

The analytic expectation of circular motion is that, the last stable 
circular orbit for a particle 
moving around a Schwarzschild black hole in a Keplerian disk is 
at $r =6M$ ($M$ is mass of black hole). 
Therefore, the gas or particles  fall into the black hole, if their
radial position  is less than $6M$. When their radial position 
 is bigger than $6M$, they should rotate in a circular orbit. Here,
this problem is simulated  and  the numerical solution is compared to 
the analytic expectations. It tests the code with sources in
the $\phi$  direction.

\begin{center}
\begin{figure}
\centerline{\epsfxsize=14.cm \epsfysize=14.cm
\epsffile{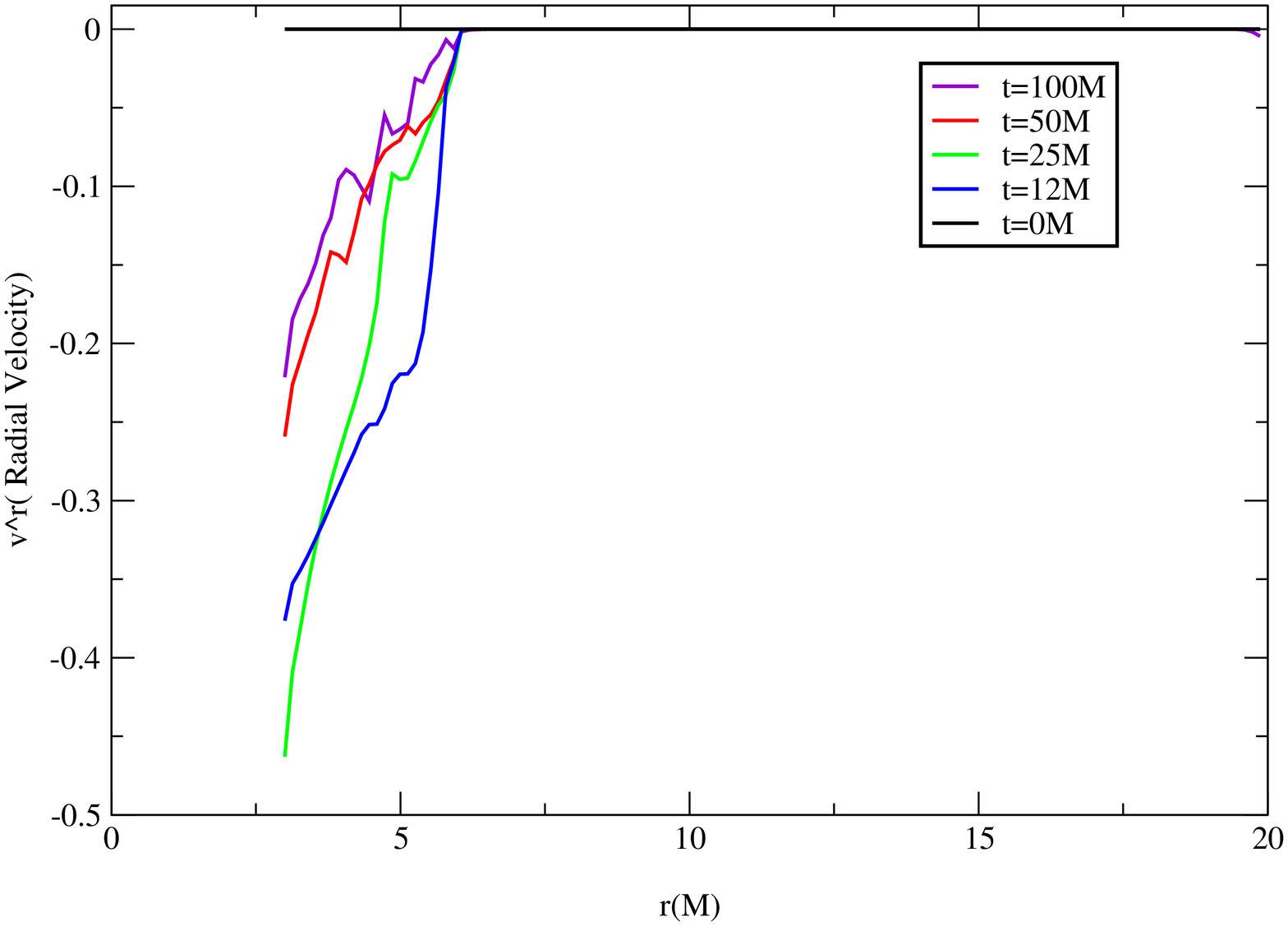}}
\caption{Radial velocity vs. $r$ is plotted. The radial velocity of fluid in the disk
stays zero for $r > 6M$, during the evolution. $r=6M$ is called the
last stable circular orbit in a Keplerian disk.}
\label{Circular motion 1}
\end{figure}
\end{center}

\begin{center}
\begin{figure}
\centerline{\epsfxsize=14.cm \epsfysize=14.cm
\epsffile{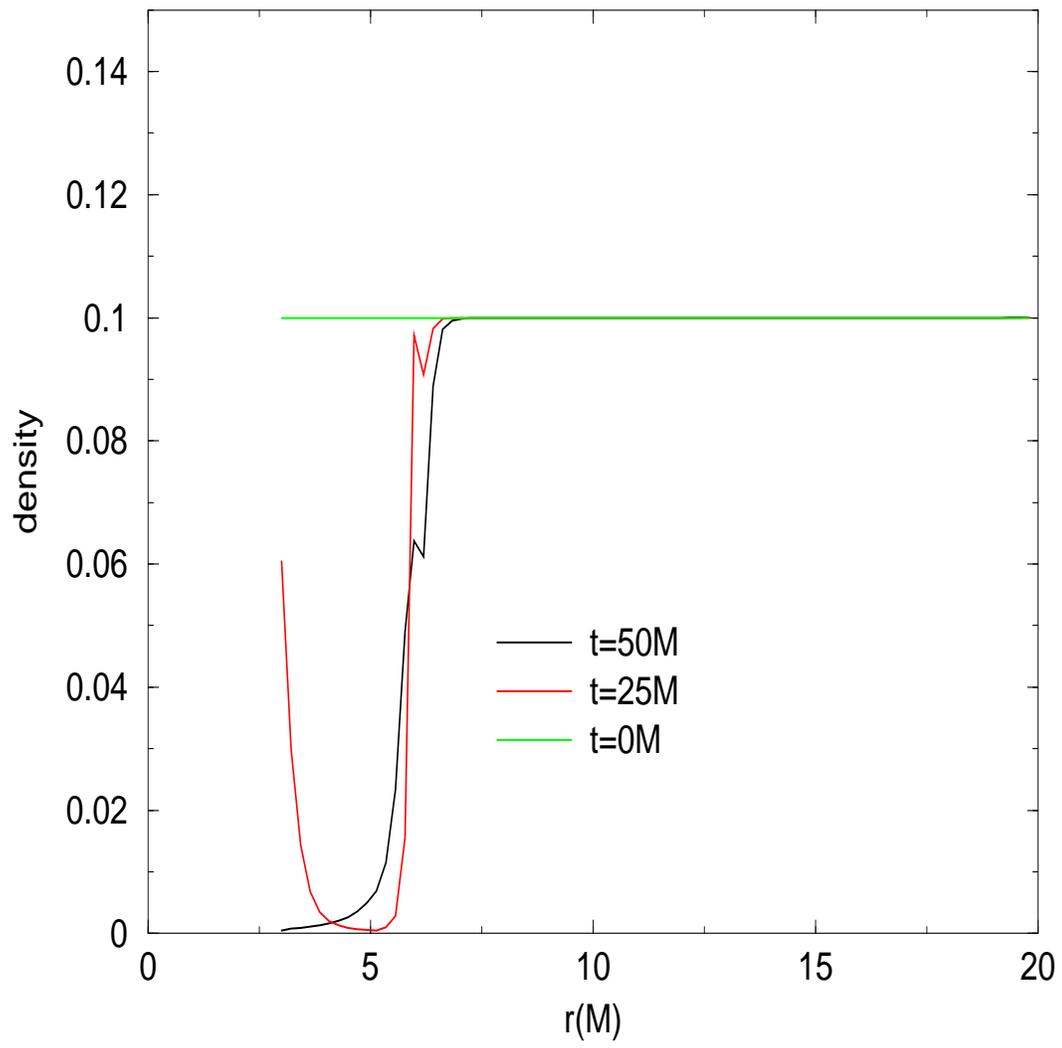}}
\caption{Density vs. $r$ is plotted. Density of the fluid is
plotted at different times using different colors. The matter falls into
the black hole while $r$ is less than $6M$.}
\label{Circular motion 2}
\end{figure}
\end{center}

In order to simulate this problem, the computational domain is chosen to be 
$3M \leq r \leq 20M$ and $0 \leq \phi \leq 2 \pi$. The
computational domain is filled with constant density and pressureless gas, 
 rotating in circular orbits with the Keplerian velocity and
zero radial velocity. In Fig. \ref{Circular motion 1}, the
radial velocities of the gas vs. radial coordinate at different
times are plotted. It is numerically observed that the gas inside the 
last stable orbit, $r=6M$,
falls into the black hole, while gas outside the last stable orbit follows
circular motion with the Keplerian velocity as it is expected analytically.
Fig. \ref{Circular motion 2} shows  the density of the fluid
 vs. radial coordinate at different times to see the 
behavior of the disk.  It is seen  that gas falls into the black
hole for  $r < 6M$. Accordingly, the numerical results from our code are
consistent with the analytic expectations.

\section{Conclusion}
\label{Conclusion}
In this paper the Eulerian type numerical solution of 
general relativistic hydrodynamic equations constructed for relativistic 
astrophysics is presented. This code is constructed for general 
spacetime metric with 
lapse function and shift vector. It is capable of solving special 
relativistic problem and general relativistic problem  coupling with the
Einstein and hydrodynamic equations. This paper explains how general 
relativistic hydrodynamic equation can be solved with AMR.

In the first part of the paper, the GRH equations and the numerical 
method are given. GRH equations are written in the conservative form and 
the numerical method and Paramesh package  is applied to them with 
some appropriate way to get working code with AMR.

Then, we have presented and tested solution of three-dimensional relativistic
hydrodynamic equations using MUSCL scheme with Marquina fluxes. The  results 
are obtained for problems involving relativistic flows, and strong shocks. 
These
problems are solved using uniform and adaptive grid. Adaptive-Mesh Refinement 
is getting one of important technique  in numerical relativity to solve the 
real astrophysical problems, such as accretion disks around the compact 
objects, and coalescing of the compact binaries. Adaptive-Mesh simulations 
pose many complex design problems, and a variety of different techniques 
that exist to solve these problems. The AMR is used  to solve 
relativistic equations and to present some results from  our numerical 
calculations. We solved the shock tube problem using different refinement 
levels to see the behavior of adaptive grid and also to compare AMR and 
uniform 
results with analytic solution. Results are converging to analytic solution 
when the number of resolutions are increased. Computing 
time used to solve AMR and uniform grid to get same accuracy are also compared.
Results show that when the number of resolutions are increased AMR uses
much less computing time than uniform grid.

Finally, The full general relativistic hydrodynamical equation are tested 
using $2-D$ test problems which are called free falling gas onto black hole
and the circular motion of particles around the black hole. The numerical 
results from these test problems agree with analytic expectation and 
converge to analytic solution when the number of resolutions are increased.

As a conclusion, AMR has permitted us to address problems of astrophysical
interest which are really hard to solve with uniform grid. Because these 
problems are dominated by local physics, this code has been designed to 
handle highly relativistic flows. Hence, it is  well suited for $2D$ and $3D$
simulations of astrophysical problems. First astrophysical results will 
be presented in a forthcoming paper for two-dimensional accretion disk 
problem around the black hole.

\begin{acknowledgments}
I would like to thank Joan M. Centrella for a  great support and 
Peter MacNeice and Bruce Fryxell for a useful discussion. This project
is supported by NSF PHY9722109 This project was carried out 
at NASA/GSFC, Laboratory of High Energy Astrophysics.
It is also supported by NASA/GSFC IR\&D. It has been performed using 
NASA super computers/T3E clusters..
\end{acknowledgments}

\vspace{2cm}

\begin{center}
{\bf References}
\end{center}

\begin{bibliography}{}

\noindent
Banyuls, F., Font, J.A., Ibanez, J.M., Marti, J.M., and
Miralles, J.A., Astrophys. J. {\bf 476}, 221 (1997).\\
Norman, M.L. and  Winker, K.-H.A., in \emph{Astrophysical
Radiation Hydrodynamics}, edited by M. L. Norman and K.-H. A. Winkler
(Reidel, Dordrecht, Holland, 1986). \\
Font, J.A., Miller, M., Suen, W.-M., Tobias, M.,
Physical Review D, {\bf 61}, 044011 (2000) \\
Donat, R., Font, J.A., Ibanez, J.M., and Marquina, A.,
J.Comput. Phys. {\bf 146}, 58 (1998).\\
Misner, C.W., Thorne, K.S., Wheeler, J.A., Gravitation\\
Schutz, B.F., in \emph{A First Course in General
Relativity}, edited by B. F. Schutz  (Cambridge University Press,
Cambridge, 1985).\\
Ibanez, J.M., and Marti, J. JCAM (1998).\\
Hirsch, C., in \emph{Numerical Computation of Internal
and External Flows}, edited by R. Gallagher and O. Zienkiewicz, Volume 2
(Wiley-Interscience, New York, 1992). \\
Quirk, J., Int. J. Numer. Met. Fl. {\bf 18}, 555-574,
(1994). \\
Shu, C.W., Osher, S.J., J.Comput. Phys. {\bf 83},
32-78 (1989). \\
Toro, E.F.,
\emph{Riemann Solver and Numerical methods for Fluid Dynamics}, 1999.\\
Bona, C., Masso, J., Seidel, E, and Walker, P., (1998),
gr-qc/9804052. \\
MacNeice, P., Olson, K.M., Mobarry, C., deFainchtein, R.
and Packer, C., Computer Physics Communications, vo.126, p.330-354,
(2000).\\
Aloy, M.A., Ibanez, J.M., Marti, J.M., Muller, E., 
Astrophysical Journal Letters, 510, 119-122 (2000)\\
Colella, P., and Woodward, P.R. J. Comput. Phys. {\bf
54}, 174-201 (1984).\\
Hawley, J.F., Smarr, L.L. \& Wilson, J.R. 1984a, ApJ, 277, 296.

\end{bibliography}

\newpage

\newpage

\newpage

\newpage

\newpage 

\newpage 

\newpage 

\newpage 

\newpage 

\newpage 

\newpage

\newpage

\newpage

\end{document}